\documentclass[structabstract]{aa}  

\usepackage{amsmath}
\usepackage{txfonts}
\usepackage{graphicx,color}
\usepackage{natbib} 
\bibpunct{(}{)}{;}{a}\ {,} 
\usepackage{multirow}
\usepackage{longtable}
\usepackage{url}

\newcommand{\kms}{km\,s$^{-1}$}
\newcommand{\ulyss}{\emph{ULySS}}

\newcommand{\eg}{\emph{e.g.}}
\newcommand{\logg}{$log$g}
\newcommand{\teff}{T$_{\rm{eff}}$}

\begin{document}
\title{Stellar population models in the UV: I. Characterisation of the New Generation Stellar Library}

\author{M. Koleva
  \inst{1,2,3,4}\thanks{Marie Curie/FWO fellow}
\and
  A. Vazdekis \inst{1,2}}
\institute{Instituto   de
  Astrof\'{\i}sica  de Canarias,  La Laguna,  E-38200  Tenerife, Spain\\
  \email{mina.koleva@gmail.com}
  \and
  Departamento de  Astrof\'{\i}sica, Universidad  de  La Laguna,
  E-38205  La  Laguna,   Tenerife,  Spain \\
  \and
  Universit\'e Lyon~1,
  Villeurbanne, F-69622, France; CRAL, Observatoire de Lyon, St. Genis
         Laval, F-69561, France ; CNRS, UMR 5574 \\ 
\and
Sterrenkundig Observatorium, Ghent University, Krijgslaan 281, S9, B-9000 Ghent,
Belgium\\
             }

   \date{Received XXX; accepted XXX}

 
  \abstract
  {
    The spectral predictions of stellar population models are not as
    accurate in the ultra-violet (UV) as in the optical wavelength
    domain. One of the reasons is the lack of high-quality stellar
    libraries.  The New Generation Stellar Library (NGSL), recently
    released, represents a significant step towards the improvement
    of this situation.  }
  {To prepare NGSL for population synthesis, we determined the 
     atmospheric parameters of its stars, we assessed the precision of the 
     wavelength calibration and characterised its intrinsic
    resolution. We also measured the Galactic extinction for each of
    the NGSL stars.}
  { For our analyses we used \ulyss, a full spectrum fitting
     package, fitting
     the NGSL spectra against the MILES interpolator.}
   {We find that the wavelength calibration is precise up to
     0.1\,px, after correcting a systematic effect in the optical
     range.  The spectral resolution varies from 3\,\AA\ in the UV to
     10\,\AA\ in the near-infrared (NIR), corresponding to a roughly constant reciprocal
     resolution $R=\lambda/\delta\lambda \approx1000$ and an instrumental
     velocity dispersion $\sigma_{ins} \approx 130$\,\kms.  We derived
     the atmospheric parameters homogeneously. The precision for the FGK stars is
     42\,K, 0.24 and 0.09 dex for \teff, \logg\ and {\rm
       [Fe/H]}, respectively. The corresponding mean errors are
     29\,K, 0.50 and 0.48 dex  for the M stars, and for the OBA stars
     they are 4.5\,percent, 0.44 and 0.18\,dex. The comparison with
     the literature shows that our results are not
     biased. 
}
{}

   \keywords{atlases -- stars: atmospheres -- fundamental parameters}
   \authorrunning{Koleva \& Vazdekis}
   \titlerunning{Stellar atmospheric parameters for NGSL}
   \maketitle


\section{Introduction}

The stellar libraries are collections of spectra that share identical
spectral coverage and resolution. They have
several important applications:  They are used as references to
classify stars and to determine their atmospheric parameters
\citep[\eg][and the reference therein]{wu2011}, as templates to
recover the line-of-sight velocity distribution of galaxies
\citep[\eg][]{ppxf}, or to calibrate photometry \citep[see][for a
review]{bessel2005}. They are one of the critical ingredients in the
stellar population synthesis \citep[\eg][]{vazdekis2010}. To produce
high-quality stellar population in the blue is the goal of this series
of papers.

The stellar libraries can be theoretical or observational.  The
theoretical libraries can, in principle, be computed for any value of
temperature, gravity, metallicity and detailed chemical composition,
and the resolution is essentially limited by the computing power
\citep[\eg][]{kurucz1979,hauschildt2003,poluxDB}.  They would be the
ideal references if they were able to reproduce the observations
accurately. In fact, the physical approximations (1D, LTE, convection,
\ldots) and the lack of complete databases of atomic and molecular
transition result in discrepancies between these stellar models and
observations \citep{martins2007,miles_params,beifiori2011}.  The
observational libraries, on the other side, have the advantage to be
assembled from real stars, but they suffer from instrumental limitations
(finite resolution) and limited atmospheric parameter coverage. Because
these libraries are constructed from Galactic stars, they are bound to the
chemical composition found in the Galaxy, and more specifically 
that of the solar neighbourhood. It is possible to combine the observed and
theoretical libraries to predict the differential effect of changing
some physical ingredients. These semi-empirical libraries were used to
extend the range of the parameter space \citep{miles_params} or to
compute the effect of a variable abundance of $\alpha$-elements in
models of stellar populations
\citep{cervantes2007,prugniel2007,walcher2009}.

The most important property of the stellar libraries is the coverage
of the atmospheric parameters, such as effective temperature, gravity,
metallicity. Recently, the detailed abundances are beginning to be considered as
important parameters, too.  Other properties to be considered are the
spectral resolution, the wavelength coverage, the flux, and the wavelength
calibration of the spectra.  Fortunately, we have optical libraries that cover
the parameter space reasonably well: ELODIE
\citep{elodie, elodie31}, CFLIB \citep{cflib}, and MILES \citep{miles}.
These empirical libraries have a fair spectral resolution
($\lambda/\Delta\lambda = R \sim 2000 - 10000$) that is compatible with the
resolution of the most widely used optical spectrographs for galactic
studies, and they have a good flux calibration \citep[except
CFLIB,][]{bruzual2007}.  They are built from normal stars in all
luminosity classes and spectral types from O to M. They cover a wide
range in metallicities ($ -3.0 \lesssim{\rm  [Fe/H]}\lesssim 1.0$).  The
stars in these three libraries, as in any other empirical library,
have the abundance pattern of the solar neighbourhood \citep[see][ for
a review]{wheeler1989}.

None of these three libraries extend further blue-ward than 3500\,\AA. The
importance of the UV as a gate to understand the physics of the stellar
systems was recognised back in the 1980s \citep[\eg][]{faber1983}. In
particular, the UV is irreplaceable to characterise the metallicity
and the star-formation history (SFH) of young populations, to study
the enhancement of $\alpha$-elements or the contribution of blue
horizontal branch stars to the integrated fluxes.  It is also of 
prime importance studying distant galaxies whose restframe UV
is observed in the optical, where the current instrumentation
is most developed. 

In a simple stellar population (SSP) the blue wavelengths are
predominantly sensitive to the hottest stars. At any age greater than
10~Myr, those are the dwarfs at the main-sequence turn-off. After
$\sim$1~Gyr of evolution, the He-burning stars may become bluer (hotter)
than the RR Lyrae pulsating stars and populate the so-called
blue-horizontal branch (BHB). Together with the blue stragglers (BS, 
low-mass, main-sequence stars with excessive blue colours) they may have
an important contribution to the integrated spectra
\citep{lee2002, cenarro2008} and mimic young populations 
\citep[\eg][]{maraston2000,koleva2008,ocvirk2009,percival2011}. 
 Our
ability to distinguish between the real young stars and these exotic
populations relies on their different contribution to the different parts
of the spectral energy distribution (SED). Thus, combining
optical and UV data can lift the degeneracy \citep{rose1984,
schiavon2004, percival2011}.

Blue-horizontal branch stars and blue stragglers are frequently
observed in Galactic clusters and were also detected in Local Group
galaxies \citep[\eg][]{mapelli2007}. The presence of the BHB and BS is
connected with some properties of the populations (\eg\ the
metallicity for the BHB) but may also be related with the environment
and with some large scale properties of the host systems.  Therefore,
the ability to distinguish these stars in integrated spectra would be
a major step toward the understanding for the genesis and evolution of
stellar systems.

It has been shown that the effects of $\alpha$-elements enhanced
partition are emphasised blue-ward, both in stars
\citep[\eg][]{cassisi2004} and in stellar populations
\citep{coelho2005}. Thus, the blue spectral range should 
provide us with more diagnostic indices to better constrain the galactic
star-formation histories \citep{serven2011}.

The first attempts to gather a UV library that covered
the MK sequence were made by \cite{IUEblue} and \cite{IUElib} with observations
from the International Ultraviolet 
Explorer\footnote{\url{http://archive.stsci.edu/iue/}} (IUE). 
This library has a resolution of 7\,\AA\ and contains 218 stars of 
essentially solar metallicity. Still, this first UV library is quite limited 
compared to its modern optical counterparts. With the new generation
stellar library (NGSL,\ \citealt{gregg2006}) the gap between the optical and
UV libraries begins to narrow.

The New Generation Spectral
Library\footnote{\url{http://archive.stsci.edu/prepds/stisngsl/}} is a
major step towards the modelling of the stellar populations in the UV.
It was observed with the Hubble Space Telescope Imaging Spectrograph
(STIS) and consists of 374 stars with metallicities between -2.0~dex
and 0.5~dex. As its optical counterparts it contains normal stars from
O to M spectral types in all luminosity classes. Its wavelength
coverage from 0.2 to 1.0~$\mu$m is the widest available amongst the
observational libraries at this resolution, though it does not go as
far in the UV as IUE and misses Ly$\alpha$. Its spectral resolution is
R$\sim$1000. The stars of the NGSL were rigorously chosen to have a
good coverage in the space of atmospheric parameters.
\citet{ngsl_params} measured the atmospheric parameters for most NGSL
stars using ATLAS9 model atmospheres \citep{castelli2004} as templates
(46 stars, i. e. $12$\,percent of the sample, miss one or more
parameters).

In order to implement this library in population models, its
characteristics have to be assessed accurately.  This is the goal of
the present paper.  In Sect. 2 we present the data, while in Sect. 3
we present our methodology. In Sect. 4 we characterise the line-spread
function (LSF) of the NGSL.  In Sect. 5 we derive the atmospheric
parameters homogeneously and compare them with the literature.  In
Sect. 6 we measure the spectroscopic Galactic extinction of the
stars. Finally, our conclusions and prospects are presented in Sect. 7.


\section{The NGSL spectra}

The NGSL stars were observed with STIS on-board HST with three
different gratings (G230LB, G430L and G750L), overlapping at
2990-3060\,\AA\ and 5500-5650\,\AA\ \citep{gregg2006}. The final
spectra cover the wavelength range from $\sim$0.2 to $\sim$1.0\,$\mu$m
(slightly different from star to star) and have a resolution of
R$\sim$1000. The flux-calibration reaches a precision of 3 percent
\citep{heap2009}. The spectra are calibrated in air wavelengths, with
sampling varying as follows: 1.373\,\AA/px ($\lambda\lambda 1675 -
3060\,\AA$ or 165\kms at 2500\,\AA), 2.744\,\AA/px ($\lambda\lambda
3060 - 5650\,\AA$ or 205\kms at 4000\,\AA), 4.878\,\AA/px
($\lambda\lambda 5650 - 10196\,\AA$ or 183\kms at 8000\,\AA). Details
about the data reduction can be found in
\url{http://archive.stsci.edu/pub/hlsp/stisngsl/aaareadme.pdf}. We
downloaded version 2 of the reduced data.

The stars were chosen to sample four metallicity groups, roughly 150
stars in each bin: {\rm [Fe/H]}$ < -1.5$; $ -1.5 <{\rm [Fe/H]}< -0.5$;
$ -0.3 < {\rm [Fe/H] }< +0.1$; $ +0.2 < {\rm [Fe/H]}$. The targeted
sample included 600 stars. Unfortunately, about 200 stars were not
observed owing to the failure of STIS in 2004. The released library
lacks some hot- and low-metallicity stars, but is well-suited to model
intermediate- and old-aged stellar populations.


\section{Methodology}
\label{sec:fitting}

We applied a full spectrum fitting approach to characterise the NGSL
spectra and to infer the stellar parameters. For this purpose we
employed the \ulyss\ package \citep{ulyss}. We followed the approach
used in \citet{wu2011}, \citet{miles_params}, and \citet{wu2011b} to
derive (i) the LSF to describe the intrinsic resolution and its
variation with wavelength, (ii) the atmospheric parameters of the
stars, and (iii) the Galactic extinction on the line-of-sight of each
star.

\subsection{Spectral fitting}

\ulyss\ performs a
parametric minimisation of the squared differences between an
observation and a linear combination of non-linear models as
\begin{align}
Obs(\lambda) = P_{n}(\lambda) \times \bigg(
    &G(v_{sys},\sigma) \nonumber \\
    &\otimes \sum_{i=0}^{i=k} W_i \,\, {\rm CMP}_i\,(a_1, a_2, ...,\lambda) 
\bigg), \label{eq:main}
\end{align}
where $Obs(\lambda)$ is the observed one-dimensional spectrum 
function of the wavelength ($\lambda$), sampled in $log\lambda$; $P_n$
is a multiplicative polynomial of degree $n$; and $G(v_{res},\sigma)$
is a Gaussian broadening function parameterised by the residual
velocity $v_{res}$, and the dispersion $\sigma$ (see the discussion in
Sect.\,\ref{sec:lsf}).  The ${\rm CMP}_i$ are $k$ non-linear functions of
any number of parameters, figuring the physical model. Their weights
$W_i$ can be constrained (to be positive in the present case).


\subsection{Applications}

Here we will use three different specific cases of
Eq.\,\ref{eq:main}.  First, to determine the broadening by comparing
the stars in common between NGSL and a reference library, we used
a single component that consists in a template spectrum (i.e. no
non-linear parameter).  Eq.\,\ref{eq:main} degenerates to
\begin{align}
\label{eq:star}
Obs(\lambda) = P_{n}(\lambda) \times G(v_{sys},\sigma) \otimes  S(\lambda), 
\end{align}
where $S(\lambda)$ is the template spectrum.

Second, to determine the broadening with respect to a theoretical
library, we used a positive linear combination of spectra taken
from a grid.
\begin{align}
\label{eq:opti}
Obs(\lambda) = P_{n}(\lambda) \times \bigg( 
    &G(v_{sys},\sigma) \nonumber \\ 
   &\otimes \sum_{i=0}^{i=k} W_i \,\, {\rm S}_i\,(\lambda) 
\bigg), 
\end{align} 
where ${\rm S}_i$ are the $k$ template spectra. The
weights $W_i$ are bound to be positive. 

Finally, we measured the atmospheric parameters of the stars using
a TGM component, as
\begin{align}
\label{eq:tgm} Obs(\lambda) = P_n \times G \otimes TGM (T_{\rm eff},
log{\rm g}, {\rm [Fe/H]}, \lambda),\end{align}
where TGM is a model spectrum, function of the effective temperature,
surface gravity and metallicity, respectively, written as \teff, \logg,
and [Fe/H].  The free parameters in the minimisation are the degree of
the polynomial, $v_{res}$, $\sigma$, \teff, \logg, and [Fe/H].

The model used for the TGM component was the MILES interpolator,
presented in \cite{miles_params}. This interpolator returns a spectrum
for any temperature, metallicity, and gravity where each wavelength
bin is computed by an interpolation over the entire reference library.
It is constructed from three different sets of polynomials for the
OBA, FGK and M type temperature ranges, and it is linearly
interpolated in overlapping regions.  Each of those sets of
polynomials are valid for a wide range of parameters, which means that this is a
global interpolation. The MILES interpolator \citep{miles_params} has
the advantage to be derivable and continuous everywhere, which makes it
suitable for non-linear minimisation as \emph{e. g.} in \ulyss.

The purpose of the polynomial is to absorb the discrepancies in the
global shape of the energy distribution between the observation and
the model, which can result from the extinction along the
line-of-sight, or uncertainties in the flux calibration.  The biases
that a prior normalisation of the observations would introduce are
minimised, because this continuum is fitted in the same time as the
parameters of the model.  If the flux calibration and shape of the
energy distribution of the model can be trusted, $P_n$ can be used to
estimate the extinction (Sect.\ref{sec:e(b-v)}). The choice of the
polynomial degree is governed by the wavelength range, the precision
of the flux calibration, the spectral resolution, and the complexity
of the fitted model.  \citet{wu2011} have shown that a degree as high
as $n = 100$ did not bias their determination of the atmospheric
parameters of CFLIB.  However, in the present case, where the spectral
resolution is lower, we found that $n > 25$ may affect the
determination of the atmospheric parameters (the polynomial competes
with the model to fit the broadest features).  We investigated the
dependence of the fitted atmospheric parameters with the degree of the
polynomial using 10 stars from each luminosity group (OBA, FGK and M).
We found that at $n = 12$ the resulting atmospheric parameters are
stable.  


\section{Line-spread function}
\label{sec:lsf}

In Eq.\,\ref{eq:main} to \ref{eq:tgm} the models are convolved with
G to match the observation. This convolution is usually meant to account
for the physical broadening of the spectrum 
when both the observation and the model have the same resolution.
This is used to measure the internal kinematics (velocity and velocity
dispersion) of galaxies \citep{tonry1979, ppxf} and requires the
logarithmic sampling of the spectrum. The same approach may be used to
measure the line-of-sight velocity and the rotational broadening of
stars.

When the two spectra have different resolutions, $G$ encompasses the
physical broadening and the relative broadening between the
observation and the model. This can be written as $G = G_{\rm phy}
\otimes LSF_{\rm rel}$, where $G_{\rm phy}$ is the physical broadening
and $LSF_{\rm rel}$ the relative LSF of the observation with respect
to the template (note that the resolution of the models should be
higher than that of the observation).  The LSF (or
relative LSF) generally depends on the wavelength, and the match of
resolution cannot be written as a convolution. However, because it
changes slowly with the wavelength, we can assume that
Eq.\,\ref{eq:main} holds in small wavelength intervals, and the
\ulyss\ analysis of the spectrum in a series of consecutive
wavelength segments will allow us to monitor the wavelength dependence
of $G$.  In addition, as in the present case, $G_{\rm phy}$ is
generally negligible compared to $LSF_{\rm rel}$, this approach will
allow us to derive the wavelength-dependent LSF.  ($G_{\rm phy}$ is
not negligible only for some fast rotating stars that we will exclude
when computing the LSF.) The first moment of the LSF (a
velocity-shift) represents the errors in the wavelength calibration
and reduction to the rest-frame velocity. The second moment
(instrumental velocity dispersion) represents the resolution.  These
two moments are likely variable throughout the library because of
slightly different observing conditions (\eg\ centring of the star in
the slit) and/or data reduction.  The dispersion relation of NGSL was
determined using the stellar lines because no calibration arc-lamp
exposures were available. This process may limit the precision of the
wavelength calibration, and result in systematic distortion of the
wavelength scale that our analysis may reveal.

The (absolute) LSF is $LSF = LSF_{rel} \otimes LSF_{ref}$, where
$LSF_{ref}$ is the LSF of the reference spectra that are known or can
be measured.  Because in the present work we consider Gaussian LSFs, the
absolute broadening can be derived by quadratically summing the
broadening of the reference spectra and the relative broadening
(returned from the optimisation algorithm).

The most straightforward and robust choice is to use high-resolution
spectra of some NGSL stars as reference to derive the LSF.
Because the spectral coverage of NGSL is wider than any other library, we will 
perform independent comparisons in the different wavelength domains. In the
optical range, we compared the stars in common with the ELODIE and
MILES libraries. To complete the LSF determination in the blue, we will
use the UVBlue theoretical grid \citep{uvblue}. Finally, we used a
grid of theoretical spectra from \citet{munari} to construct the LSF
over most of the range, except for the first 500\,\AA\  at the blue end. The
different analyses were cross-checked in their wide overlapping
spectral regions. 

\subsection{ELODIE and MILES libraries}

To check the wavelength calibration of the NGSL stars in the optical,
we compared the stars in common between NGSL and MILES\footnote{For
  the interpolator based on MILES, \cite{miles_params} used the first
  official version of the library (v.9.0), while for the LSF
  comparison we used the new 9.1 version, where the wavelength
  calibration of some stars was re-done.}\citep{miles,falcon2011} or
ELODIE \citep{elodie,elodie31} libraries. In that case,
Eq.\,\ref{eq:star} applies. Since the same objects were observed in
the two libraries, the only difference between the spectra should be
the instrumental broadening. We proceeded as follow: first we fitted
the `observed' against the `template' star to clean residual spikes in
the spectrum. Second, we fitted a Gaussian broadening in segments of
400\,\AA\ separated by 300\,\AA\ (so that the consecutive segments
overlap by 100\,\AA\ at both ends). Third, we derived the absolute LSF
by adding quadratically the LSF of the reference library. For ELODIE
we took a FWHM=0.58\,\AA\ and for MILES a FWHM=2.5\,\AA\
\citep{miles_params,falcon2011}. Finally, the results were averaged to
produce a {\it mean} LSF. The individual measurement outliers, from
either particular stars (imprecise reduction to the rest-frame or fast
rotators) or poor fits (low S/N) were rejected at this step using at
most five iterations of a $3-\sigma$ clipping (with the IDL function
{\sc meanclip} from the Astrolib library\footnote{Note
  that in some cases the distribution is very skewed, therefore even with
  clipping the mean may not coincide with the peak of the
  distribution.}). We also rejected the measurements where the broadening varied
significantly between two successive segments.

\begin{figure}
\includegraphics[bb=44 360 337 586, width=0.47\textwidth]{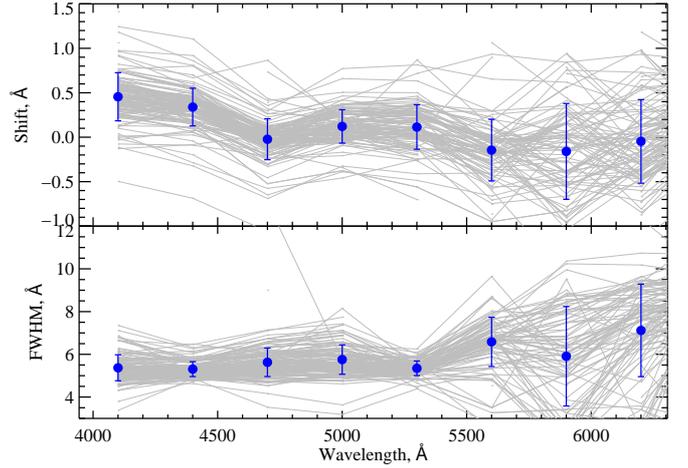}
\caption{Relative LSF for the 127 NGSL stars in common with the ELODIE
  library. The panels show the shift in wavelength and the resolution
  (FWHM) as a fucntion of wavelength. The obtained LSFs for the
  individual stars are plotted in thin grey lines. The mean values and
  their standard deviations are plotted in blue.  }
  \label{fig:lsf1}
\end{figure}

\begin{figure}
\includegraphics[width=0.47\textwidth]{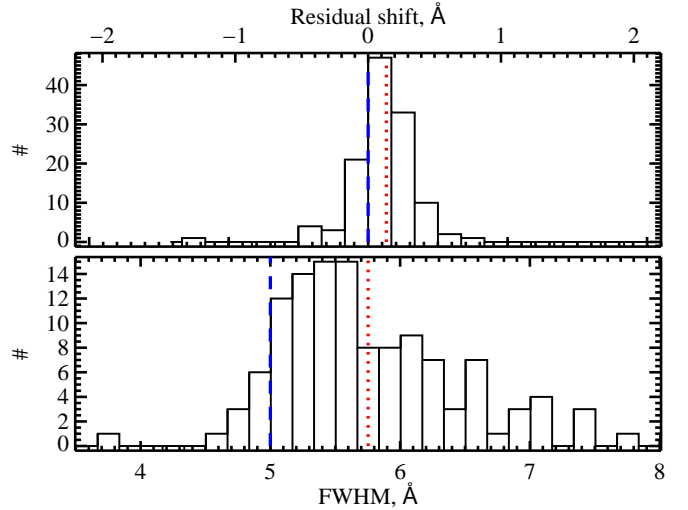}
\caption{Histograms corresponding to the distributions of the
    wavelength shift and the resolution at $\sim$5000\,\AA\ for the
    127 NGSL stars in common with ELODIE. The blue dashed vertical
    lines mark the zero velocity and the expected dispersion of
    5\,\AA\ , while the red dotted lines are the mean residual shift
    and the mean FWHM (top and bottom respectively).  }
  \label{fig:lsf2}
\end{figure}

The results for 127 stars in common with ELODIE are plotted on
Fig.\ref{fig:lsf1}, \ref{fig:lsf2}. The results from the 137 MILES comparisons are
similar and consistent, though with a wider (by $\sim$25\,\kms)
spread in velocity, consistent with the internal spread of MILES stars
found in \citet[][12\,\kms]{miles_params}. The residual shifts and the
Gaussian widths varies from star to star. At 5000\,\AA\ the internal
dispersion of the residual shift for the ELODIE comparison is
0.32\,\AA\ (equivalent to an internal dispersion of 19\,\kms, or
$\sim$0.1 pixel), and the standard deviation of the FWHM broadening is
0.75\,\AA\ (or 45 \kms).  The spread of the residual shift is similar
to the the typical uncertainties from data reduction (usually
10\,percent or lower).  The variance of the FWHM width is partly
physical (i.e.  rotation) and partly instrumental (star-to-star
difference of resolution).

\subsection{UVBlue theoretical grid}
The UVBLue library \citep{uvblue} covers the wavelength range from
87~nm to 470~nm at $R=50000$. Its parameter coverage reaches from
\teff\ = 3000 to 50 000~K, $log$g = 0.0 to 5.0 with steps of 0.5 dex,
and {\rm [Fe/H]}= -2.0 to 0.5 dex, computed at 7 nodes with a solar
mix \citep{anders1989}. The final grid consists of 1770 models with
local thermodynamic equilibrium (LTE). These were computed using the
updated list of atomic transition given by \cite{kurucz1992}, adding
all diatomic molecular lines except TiO. However, the latest molecule
transitions are only prominent in late-type stars with \teff\, bellow
4000 K, which in turn do not have significant UV flux. The
microturbulent velocity was fixed ($\xi$ = 2\,\kms, default for the
Kurucz models). This value is expected to vary in the different
spectroscopic classes from 1.5 to 10\,\kms, but the effects are
expected to be negligible at low resolution (as in the present case).
We downloaded the $R=10000$ version available
online\footnote{\url{http://www.inaoep.mx/~modelos/uvblue/uvblue.html}},
calibrated in air wavelength. We measured its intrinsic resolution
using the solar spectrum from the BASS2000
database\footnote{\url{http://bass2000.obspm.fr/solar_spect.php}} and
found it to be consistent with the value given by the authors.

Each of the NGSL stars was fitted against a positive linear
combination of UVBlue spectra according to Eq.\,\ref{eq:opti}.  For
each NGSL stellar spectrum the comparison was made with the eight
UVBlue spectra whose parameters surround the values from
Table\,\ref{table:params} (determined in Sect.\,\ref{sec:param}).  We
determined the weights of the different UVBlue spectra from a fit over
the whole wavelength range, and we used this combination to analyse
the individual segments. This first fit of the whole spectral range
was also used to clip the spikes.  We performed the LSF analysis again
in the 400\,\AA\ segments and we averaged the individual LSFs.

While fitting the LSF we noticed that the region between 2400 and
3000\,\AA\ is poorly matched by the theoretical spectra. This
discrepancy was already pointed by the authors of UVBlue
\citep{uvblue}, stating that the simulated FGK stars fail to reproduce
important prominent metallic features as the Fe{\sc ii} blend at
2400\,\AA, Fe{\sc i}/Si{\sc i} blend at about 2500\,\AA, Mg{\sc
  ii} doublet at 2800\,\AA\, Mg{\sc i} line at 2852\,\AA\ and the Mg
break at $\sim$\,2600\,\AA. We find that the match is also poor for
the OBA and M spectral types (although the low flux of the M stars in
the blue reduces the significance of the comparisons because of the
S/N limitation). However, even though that many prominent features are
misfitted, there is enough information in the other lines to constrain
the LSF, but of course with a lower precision. Hence, the LSF
determined in this region needs to be taken {\it cum grano salis}.

\subsection{Munari et al. theoretical grid}

To assess the LSF in the NIR part of the NGSL stars, we downloaded the
1\,\AA/px, scaled solar version of the Munari
library\footnote{\url{http://archives.pd.astro.it/2500-10500/}}
\citep[$\lambda\lambda = 2500-10500$\AA,][]{munari}. The library of
51288 spectra was produced using Kurucz models covering the space of
atmospheric parameters as follows: $3500 \le $~\teff~$\le 47500$\,K,
$0.0 \le $~\logg~$\le 5.0$, $-2.5\le [M/H] \le 0.5$, $[\alpha/Fe] =
0.0, +0.4$, $\xi =1,2,4$\,\kms, $0 \le V_{rot} \le 500$\,\kms. They
used the updated list of atomic transition by \cite{kurucz1992},
adding some molecular lines, including TiO for stars with \teff~$<
5000$\,K. The authors compared the colours and temperatures of their
predictions to other synthetic libraries and real stars, noticing that
their models fail to reproduce the very red colour observed in
low-temperature stars.

We measured its intrinsic resolution using the solar spectrum from the
BASS2000 database and found it to be FWHM = 2.1\,\AA\ throughout the
full wavelength range. Again, the spectra are
air-wavelength-calibrated.  The Munari grid was generated at high
resolution, then convolved with a Gaussian to lower resolution and
finally rebinned to pixels of half the FWHM of this Gaussian.  Since
the rebinning also implies a convolution by a top-hat function of
1\,pixel, the final broadening is slightly larger than the convolving
Gaussian.  We performed the LSF analyses over the full wavelength
range in the same way as for UVBlue, following Eq.\,\ref{eq:opti}.

\subsection{Corrected wavelength calibration and adopted LSF}

\begin{figure}
\includegraphics[bb=44 360 337 558, width=0.49\textwidth]{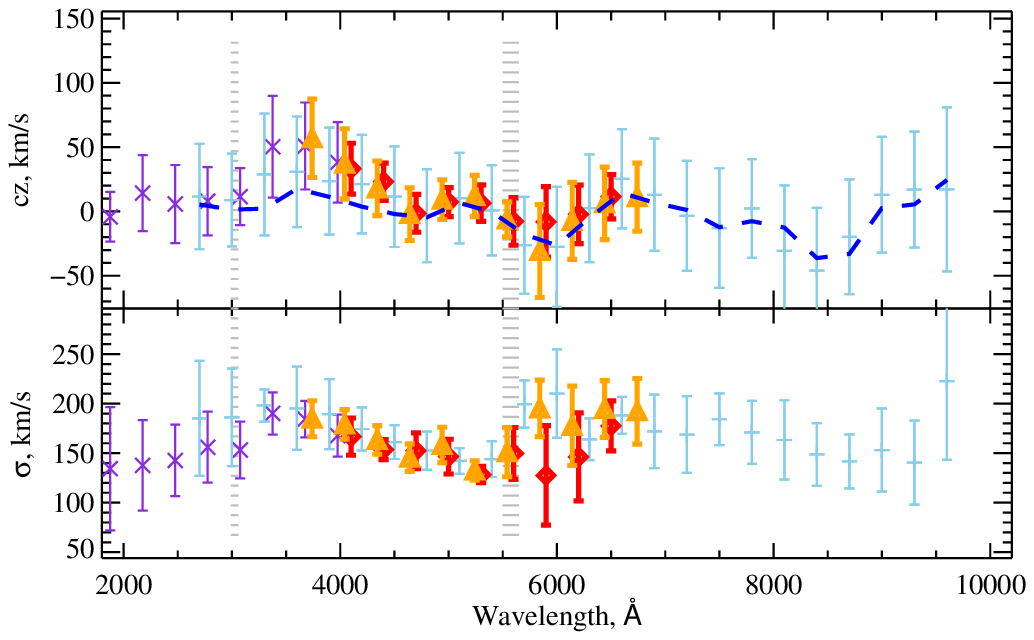}
\includegraphics[bb=44 360 337 558, width=0.49\textwidth]{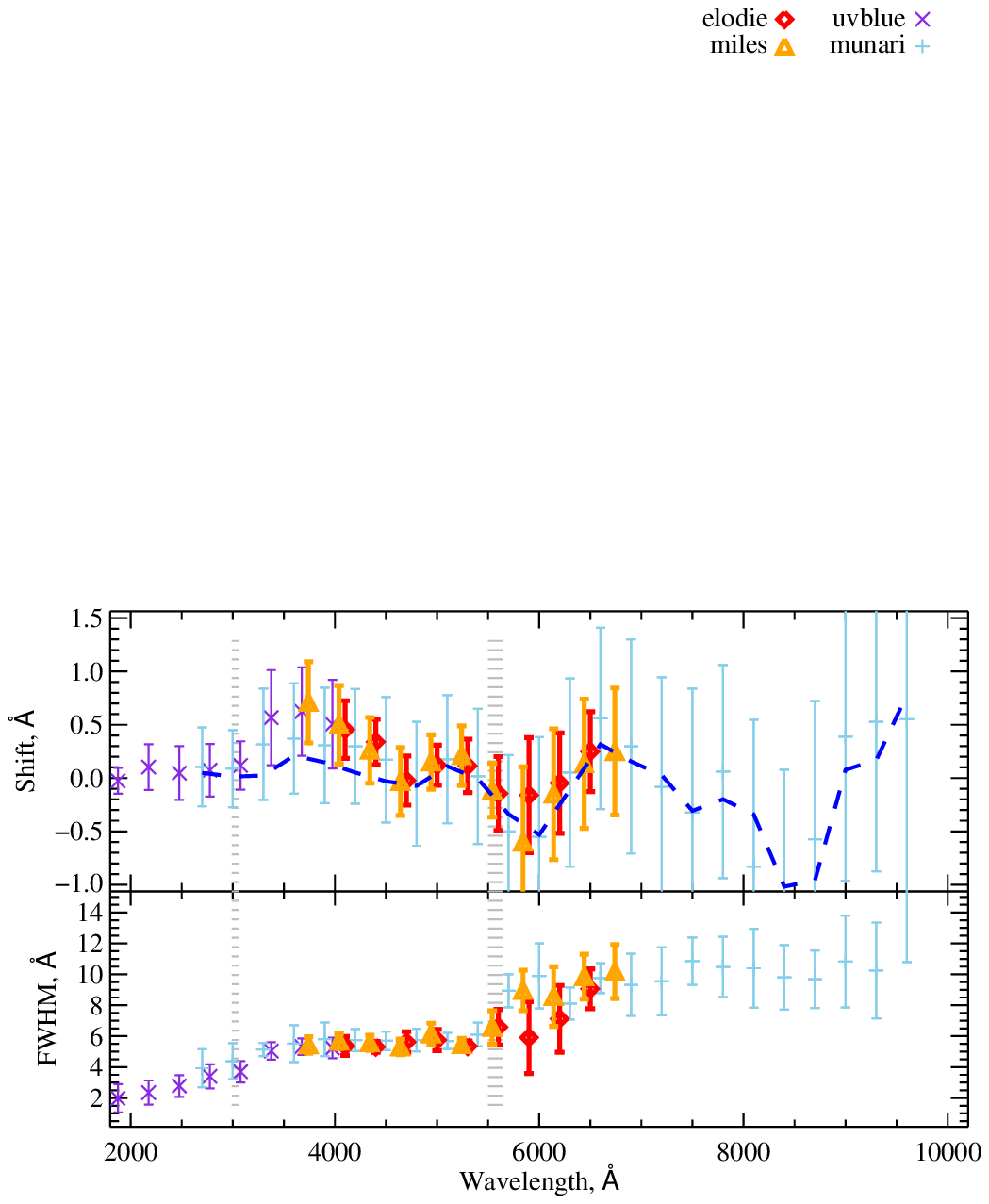}
\caption{Global line-spread function of NGSL.  The upper panels show
  the residual velocity and instrumental velocity dispersion in
  \kms. The bottom panels show the corresponding wavelength shift and
  resolution in \AA. The LSFs obtained with the aid of the various
  reference libraries are overplotted with the symbols according to
  the legend in the first panel. The error bars represent the standard
  deviations from the distribution of the LSF obtained with the
  individual stars.  The regions where the three segments of NGSL
  overlap are marked with grey horizontal lines. The blue dashed line
  shows the applied correction to the velocity shift. (MILES and
  ELODIE libraries have different starting points, therefore their results
  appear slightly shifted.) }
\label{fig:mean_lsf}
\end{figure}

The LSFs obtained with the four reference libraries are represented in
Fig.\,\ref{fig:mean_lsf}. The results from these comparisons are fully
consistent.

The FWHM of the LSF is varying from 3\,\AA\  at the UV end
to 5\,\AA\  at 5000\,\AA\ , and to 10\,\AA\  
at the NIR end.  It is roughly constant over each segment
corresponding to the different gratings, and the discontinuities
between the three gratings, at 3060 and 5650\,\AA\ , are clear.

The residual shifts for the UV and red gratings do not significantly
depend on the wavelength and are small: 10 and 0\,\kms. Our analysis
reveals a defect of the wavelength calibration of the green segment
(G430L grating). We used a simple linear relation to correct it:
$\lambda_{cor} = \lambda - 0.7 (5650 - \lambda) / (5650 - 3060) $,
for $3060 < \lambda < 5650$ \AA, where $\lambda$ is the original
wavelength in \AA\ and $\lambda_{cor}$ the corrected wavelength.  This
first-order correction of the dispersion relation is derived from the
drift of the residual wavelength shift seen on
Fig.\,\ref{fig:mean_lsf}.  This correction can easily be applied to
the wavelength array when the NGSL FITS files are read.

We averaged the four LSFs and restored the discontinuities, which were
smoothed by our analysis in 400\,\AA\ segments by extrapolating the
trend seen for each grating towards the overlap regions.


\section{Atmospheric parameters}
\label{sec:param}

We determined the atmospheric parameters of the NGSL stars by
fitting the spectra against a reference spectrum of given \teff,
\logg\ 
and {\rm [Fe/H]} . We compared the parameters with various
previous studies to assess their reliability and precision.

\subsection{Measurement of the parameters}
\label{subsec:measure_param}

We determined the atmospheric parameters of NGSL with \ulyss\ as in
\citet{wu2011}, \citet{miles_params}, and \citet{wu2011b}. The fit is
performed according to Eq.\,\ref{eq:tgm} over the wavelength range
3500 to 7500\,\AA\ (the wavelength range of the MILES stellar library).
The spectra were logarithmically rebined to pixels corresponding to
100\,\kms. We used the LSF
derived in the previous section as described in
\citet{ulyss}. Because of the variation of the LSF throughout the
library, the convolution by $G$ in Eq.\,\ref{eq:tgm} was
maintained. To prevent that the model adapted to
the mean NGSL resolution becomes broader than a given spectrum, we
biased the injected LSF by 100 \kms\ (i.e. we subtracted 100 \kms in
quadrature from $\sigma_{ins}(\lambda)$).  We used a 12$^{th}$ degree
multiplicative polynomial to absorb any continuum mismatches between
the model and the observations (Sect.\,\ref{sec:fitting}).
 
\ulyss\ performs a local minimisation starting from a guess point in the
parameter space. Thus, the solution may be trapped in a local minimum.
To avoid this and reject local minima, 
we repeated the minimisation from multiple guesses, sampling the
parameters spaces at the following nodes:
\teff\ in $[3500, 4000, 5600, 7000,
10000, 18000, 30000]$, {\rm  [Fe/H]}\ in $[-1.7, -0.3, 0.5]$ and \logg\ 
in $[1.8, 3.8]$. 

\begin{figure}
\includegraphics[width=0.50\textwidth]{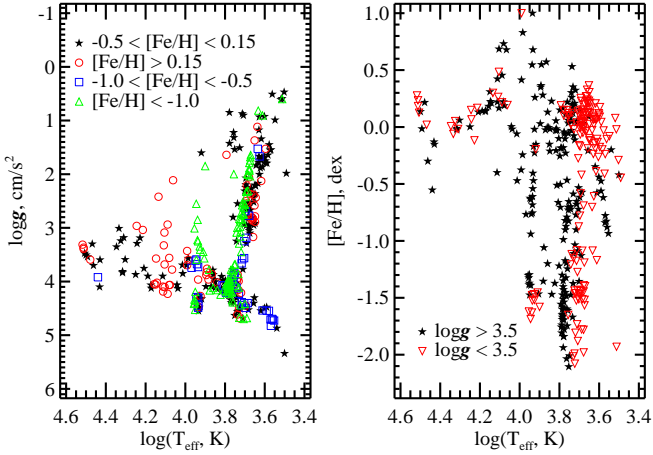}
\caption{Resulting stellar parameter coverage for the 374 stars of
  the NGSL. The left panel shows their distribution in the \teff\ - \logg\
  plane. These stars were separated into four different metallicity bins
  according to the legend.  In the right panel we plot the dwarf and
  giant distribution in the \teff\ -{\rm [Fe/H]}\ plane.  }
\label{fig:params}
\end{figure}

 Previous studies, in particular \cite{wu2011}, have shown that while
this method is highly reliable for FGK stars, special care has to
be taken for OBA and M stars. There are various reasons for this lower
reliability: First, the reference library is more scarcely populated
in these regions of the parameters space; second, high-precisions
measurements of the atmospheric parameters of these stars are not as
abundant as for FGK stars; finally, the determination of the
parameters of those stars is more complex for physical
reasons. Therefore, we systematically checked our determinations
against recent literature for OBA and M stars as well as for cases
that we found in disagreement with other studies.  For 30 stars
(8\,percent of the library) we adopted measurements from the
literature.  Our final list of parameters are given in
Table\,\ref{table:params}. The distribution of the stars in \teff\ -
\logg\ and \teff\ - {\rm [Fe/H]} planes is presented in Fig.\,\ref{fig:params}.


\subsection{Comparison with the literature}
\label{sec:comp}

We compared our measured parameters with those from four previous
studies by  
\cite{heap2009}, \cite{wu2011}, \cite{miles_params}, and \cite{soubiran2010}.  

\cite{heap2009} derived preliminary, main atmospheric parameters of the NGSL
(version 1).  They derived
\teff\ and the metallicity by fitting the stellar spectra to
models of Munari \citep{munari}. The gravity was obtained from the star position on
the HR diagram using the Hipparcos parallax for the distances.  The
comparison of their results with the \cite{valenti2005}
catalogue of cool stars shows some deviations for the \logg, which are
not unexpected because the atmospheric parameters are
coupled, and deriving them separately may lead to biases.

The determination of the CFLIB and MILES parameters
\citep{wu2011,miles_params} was performed as in the present paper.  This
method is highly reliable for intermediate spectral types, but it is
not as accurate for the extremes of the HR diagram, as discussed in
Section\,\ref{subsec:measure_param}.  There are, however, two main
differences that may affect the comparison. On one hand, the
presently analysed library has a much lower resolution than that
of CFLIB (R$\sim$5000) and MILES (R$\sim$2000). The information from
the weak lines accordingly is blended, which is particularly important
when analysing earlier stellar types. On the other hand, the MILES
interpolator, performs better than ELODIE (3.1 or 3.2) in some regions
of the HR diagram (in particular for blue-horizontal branch stars,
\citealt{miles_params}). Hence, stars in this regions will possibly
have a more precise determination of their atmospheric parameters.

Finally, the latest version of the PASTEL database
\citep{soubiran2010} is a compilation of previously published
 stellar atmospheric parameters. It continues series of publications
 by Cayrel and collaborators \citep{cayrel2001}. Most of the
 measurements were made using high spectral resolution and high
 signal-to-noise data, though inhomogeneous. When there were multiple
 measurements of the same parameter of a given star available, we took
 the mean value.

The statistics of the comparisons was made without the outliers and
it is presented in Table\,\ref{table:compa}. The corresponding figures
with the four comparison are given in Appendix\,\ref{app:comparison}.

\begin{table*}
\label{table:compa} 
\caption{Comparison of the atmospheric parameters with other studies.}
\begin{tabular}{l|rr|rrr|rrr|rrr}
  \hline\hline
  Comparison & &N\tablefootmark{a} & \multicolumn{3}{c}{$T_{{\rm eff}}$\tablefootmark{b}}&  \multicolumn{3}{c}{log g (cm s$^{-2}$)} &  \multicolumn{3}{c}{[Fe/H] (dex)}\\
  &&&\multicolumn{1}{c}{$\Delta$}&\multicolumn{1}{c}{$\sigma$}&\multicolumn{1}{c}{$\nabla$}&\multicolumn{1}{c}{$\Delta$}&\multicolumn{1}{c}{$\sigma$}&\multicolumn{1}{c}{$\nabla$}&\multicolumn{1}{c}{$\Delta$}&\multicolumn{1}{c}{$\sigma$}&\multicolumn{1}{c}{$\nabla$}\\
  \hline
  \multirow{3}{*}{NGSL} 
 &        OBA &  59 &   4 &   3 &  0.87 &  0.12 &  0.47 &  0.45 & -0.10 &  0.60 &  0.56\\
 &        FGK & 236 &   4 & 171 &  1.08 &  0.22 &  0.42 &  1.18 &  0.09 &  0.34 &  0.94\\
 &          M &   6 & -42 & 120 &  1.14 &  0.32 &  0.55 &  1.32 & -0.35 &  0.40 & -0.05\\
\hline
\multirow{3}{*}{MILES} 
 &        OBA &  20 &   6 &   6 &  0.96 &  0.11 &  0.45 &  0.87 & -0.02 &  0.19 &  0.99\\
 &        FGK &  55 &  26 & 115 &  1.01 &  0.03 &  0.28 &  1.10 & -0.00 &  0.11 &  1.01\\
 &          M &   4 &  41 &  37 &  1.02 &  0.13 &  0.14 &  0.83 &  0.08 &  0.19 &  0.06\\
\hline
\multirow{3}{*}{CFLIB} 
 &        OBA &  32 &   5 &   3 &  1.04 &  0.05 &  0.30 &  0.93 &  0.03 &  0.22 &  1.03\\
 &        FGK &  80 &  43 & 111 &  1.00 & -0.00 &  0.20 &  1.04 &  0.04 &  0.11 &  1.01\\
 &          M &   5 &  37 &  23 &  0.98 &  0.46 &  0.69 &  0.24 &  0.17 &  0.29 &  1.00\\
\hline
\multirow{3}{*}{PASTEL} 
 &        OBA &  57 &   7 &   6 &  1.06 &  0.06 &  0.43 &  0.65 & -0.00 &  0.63 &  0.90\\
 &        FGK & 176 &  69 & 196 &  1.00 &  0.04 &  0.42 &  1.09 &  0.05 &  0.19 &  1.03\\
 &          M &   5 & -73 & 246 &  0.51 & -0.13 &  0.19 &  1.02 & -0.26 &  0.33 &  0.60\\

  \hline
\end{tabular}
\tablefoot{
  For each parameter the $\Delta$ column gives the mean difference
  `this work' $-$ `reference', $\sigma$ the dispersion between the two
  series and $\nabla$ the slope from the linear fit. The three lines are for
the OBA ($T_{{\rm eff}} > 8000 K$), FGK  ($4000 < $\teff$ \leq 8000 K$) and M 
($T_{{\rm eff}} \leq 4000 K$) spectroscopic types, respectively. 
The statistics were computed discarding the outliers.}\\
\tablefoottext{a}{Number of compared spectra}
\tablefoottext{b}{The
$\Delta$ and $\sigma$ of \teff\ are in K, except for the OBA stars, where
these statistics are given in percent.}
\end{table*}

The FGK stars agree well with the results of the other
authors. The biases are within the errors of the individual parameters
and are not significant.

The linear fit of the comparison has slopes ($\nabla$) close to unity
in almost all cases.  The dispersions from the comparisons are higher
than in \cite{miles_params}, revealing a lower precision, which is
probably caused by the lower spectral resolution.

The strongest discrepancy for the FGK spectral classes appears when
comparing our \logg\ values with that derived by
\citet{heap2009} from the same data (see
Fig.\,\ref{fig:compNGSL}). Our gravities are up to one dex higher for the
dwarfs. This is consistent with the bias found in \cite{heap2009} by
comparing to \cite{valenti2005}.  
The discretisation of the measurements of \cite{heap2009} on the nodes
of the model's grid is also apparent in the figure. This is because 
the authors did not interpolate between the templates.

The automatic determination of the parameters is more difficult for
OBA and M stars. At this low resolution only few lines are present for
early-type stars, and the parameters are less constrained. In
addition, the profiles of the lines are also affected by rotation (and
inclination of the rotation axis on the line-of-sight) and some stars
display emission lines.  For the late types, on the other hand, the
spectra are dominated by broad molecular features and the individual
narrow band is lost because of the low resolution. Despite these
difficulties, the comparison with previous studies revealed no
trend or bias. 

To summarise, our results are consistent with those of authors for
all spectral types. The offsets in
\teff\ vary from 4 to 69\,K in the FGK stars, depending on the comparison study
with a dispersion of $\sim$ 150 K; in \logg\ the offset found in
comparison with \cite{heap2009} is 0.22, while there is no offset with
the other references, the typical dispersion is about 0.35\,dex; the
shift in metallicity is negligible and the dispersions are between 0.11 to
0.34\,dex. For the OBA spectral types we find $ \Delta(T_{eff}, K)
\sim 5 $\%, $ 0.05 < \Delta(log$g$, cm/s^2) < 0.12 $, with a dispersion $\sim
0.47$ and $ -0.02 < \Delta [Fe/H], dex < 0.10 $, varying between $0.2$
and $0.6$\,dex. There are too few M stars in common between the
different data sets to make any statistical analyses.

\subsection{Error estimation}

The errors returned by \ulyss\ are computed from the covariance matrix.
They underestimate the real precision because (i) the parameters
are not fully independent (for example there is the well-known
degeneracy between \teff\ and \logg), (ii) the fits are not perfect
(there are some mismatches caused by non-solar
abundances or particularities), and (iii) the errors in the data are not accurately
known.  Therefore, we first determined an upper limit to the internal
error by forcing $\chi^2 = 1$, and we estimated the external error by
rescaling the internal error as in \citet{wu2011}.

To estimate the external errors we used the statistics of the
comparison between our determination and \citet[see
Sect.\,\ref{sec:comp}]{miles_params}.  We chose this reference for
the comparison because it is homogeneous and reasonably accurate.  The
external errors were determined in Prugniel et al., therefore we
subtracted them quadratically from the dispersion of the comparison and
obtained an estimate of the mean external error. From this mean external
error we derived the rescaling factor. For a given spectral type and
stellar parameter, the rescaling factor is computed as
\begin{align}
\xi = \frac{\sqrt{\sigma_{tot}^2 - \bar{\sigma}_{mil}^2}}{\bar{\sigma}_{int}},
\label{eq:rescfact}
\end{align}

where $\sigma_{tot}$ is the residual dispersion from corresponding
comparison in Table\,\ref{table:compa}, $\bar{\sigma}_{mil}$ the
external errors reported in \citet{miles_params}, and
$\bar{\sigma_{int}}$ the mean internal error from the present fits.
The final corrected error for each of the stars is $\sigma'_i = \xi
\times \sigma_i $.

The correction coefficients are about 2.5 for the FGK stars. For the
early spectroscopic classes they vary from 2.5 (for metallicity) to
6.0 (for gravity). We do not have sufficiently high statistics to compute
these coefficients for the M class. However, we consider that the
external errors are roughly the same as for the OBA stars and as in
\citet{miles_params}. Therefore, we used their coefficients to
correct the errors of M stars. The median precision of the derived
parameters is 42 K for \teff, 0.24 dex in \logg\ and 0.09 dex in
[Fe/H] for the FGK class. For the OBA stars they are 4.5\,percent,
0.44 dex and 0.18 dex, and for the M stars 29\,K 0.50\,dex and
0.48\,dex for temperature, gravity, and metallicity, respectively. The
precisions are lower than those obtained by \citet{miles_params},
probably because of the lower spectral resolution.  


\section{Galactic extinction}
\label{sec:e(b-v)}

The Galactic extinction may be determined from photometry
\citep[\eg][]{neckel1980} or using a Galactic model
\citep{chen1998,hakkila1997}.  The first method requires (i)
accurate photometry and (ii) a good knowledge of the intrinsic
spectral energy distribution of the stars (i.e. a precise spectral
classification).  A wrong estimate of the metallicity will immediately
translate into an error on the Galactic extinction.  The second
method requires knowledge of the direction and distance to the
star and an adequate model of the Galaxy. For example the
\cite{chen1998} model is a simple geometric representation of the
galaxy scaled to the ``total'' extinction provided by the Schlegel
maps. Alternatively the \cite{hakkila1997} model is calibrated on
empirically measured extinctions in several directions in the galaxies.
These models are generally acceptable in low-extinction regions
(i.e. Galactic latitude $|b| > 10^\circ$), but they are less reliable 
in high-extinction regions.
 
An alternative to these two methods is to directly measure the
extinction on the NGSL spectra.  The NGSL spectra were flux-calibrated
with a precision of 2-3\,percent \citep{heap2009}.  Therefore, $P_n$ mixes
information about the uncertainty of the flux calibration and 
the Galactic extinction.  Hence, we can assume to first approximation
that $P_n$ (Eq.\,\ref{eq:tgm}) derived in Sect.\,\ref{sec:param}
corresponds to the extinction curve.  We therefore fitted $P_n$
against the Galactic extinction law from \citet{fitz1999}. The
precision on the derived E(B-V) colour excess depends on the precision
on (i) the flux calibration, (ii) the atmospheric parameters, and
(iii) the best-fitted template.  The precision of the ELODIE and MILES
interpolators were discussed in \cite{miles_params} and were found to be
accurate to 1-2\,percent.  

The extinction law, $A(\lambda)/A_V$ (normalised to the V band)
depends on the line-of-sight. It can be parameterised with $R_V = A_V
/ E(B-V)$ (the ratio between the extinction in the V and the B-V colour
excess), which have a `mean' value of 3.1, but it varies between
2.3 and 5.3 \citep{cardelli1989, fitz1999}. The extinction is almost
independent of $R_V$ in the red, but is strongly dependent on the
wavelength in the blue and UV.  Adopting the \cite{fitz1999}
extinction law, we were able to fit simultaneously $A_V$ and $R_V$ over
almost the full wavelength range by comparing the observed spectra to
the Munari best fit. However, it is known that the SED of the
theoretical spectra might not provide a good match to the empirical
counterpart, particularly in the shortest wavelengths. For this
reason, we preferred to compare the observations to the best-fitted
interpolated MILES spectrum obtained in Sect.\,\ref{sec:param}.

Fitting the extinction using MILES restricts the wavelength range to
the optical domain, and therefore the correction of the whole spectrum
requires extrapolations. The extrapolation towards the infrared should
be safe because the extinction law is uniform in any line-of-sight
(and the extinction is low), and the quality of the corrected spectrum
will be essentially limited by the precision of the original flux
calibration. However, the extrapolation towards the UV can be more
hazardous because the determination of ${\rm R_V}$ will only rely on
the blue end of the MILES spectra. Any error on the extinction law
would be amplified in the UV. For this reason, we preferred here to adopt
$R_V = 3.1$ and we fitted only $A_V$ over the wavelength range
3500-7500\,\AA. 

To test the reliability of these determinations of the extinction, we
compared them with the predictions of the \citet{chen1998} Galactic extinction 
model for the stars with parallaxes known from
Hipparcos\footnote{\url{http://www.rssd.esa.int/index.php?project=HIPPARCOS&page=index}}.
This comparison is acceptable with a slope of 0.85.
Our values of $A_V$ are listed in Table\,\ref{table:params}.


\section{Conclusions}

We have fully characterised the NGSL for its implementation in stellar
population synthesis modelling. We used \ulyss, a full spectrum fitting
package.
We found that the line-spread function of the
stellar spectra of this library vary from 3\,\AA\ in the UV to 10\,\AA\ (FWHM) in
the near IR. The instrumental velocity dispersion is virtually constant within
the whole spectral range covered by the library, at $\sigma_{ins} \approx 130
$\,\kms.  The wavelength calibration is accurate to 0.1\,px
(0.32\,\AA\ at 5000\,\AA).  We
measured the atmospheric parameters of the stars using the \ulyss\ 
package and the MILES interpolator \citet{miles_params}. By comparing
the results to previous studies we found that the precisions for
the FGK stars are
42\,K, 0.24 and 0.09 dex for \teff, \logg\ and {\rm
  [Fe/H]},  respectively. For the M stars, the corresponding mean errors are 29\,K, 0.50 and
0.48 dex, and for the OBA 4.5\,percent, 0.44 and 0.18\,dex. Finally,
we measured the Galactic extinction for each star by directly comparing
the spectra to the interpolated MILES spectra.

The NGSL library is a major step towards the accurate modelling of
stellar populations over a wide wavelength range. In the second paper
of this series we make use of the NGSL and the results of this work to
expand the spectral coverage of our stellar population models
\citep{vazdekis2010}.

\begin{acknowledgements}
  This work has been supported by the Programa 
  Nacional de Astronom\'{\i}a y Astrof\'{\i}sica of the Spanish Ministry 
  of Science and Innovation under grant \emph{AYA2010-21322-C03-02}.
  MK thanks CRAL, Observatoire de Lyon, Universit\'{e} Claude
  Bernard for an Invited Professorship and Ph. Prugniel for the
  useful comments. She is a postdoctoral fellow of the Fund for
  Scientific Research-Flanders, Belgium (FWO11/PDO/147) and Marie
  Curie (Grant PIEF-GA-2010-271780). We would like to thank the
  referee for her/his helpful report. \end{acknowledgements}

\bibliographystyle{aa} 
\bibliography{ngsl}   

\Online

\begin{appendix} 

\section{Atmospheric parameters}

Here we list the adopted parameters of the 367 NGSL stars
(Table\,\ref{table:params}), together with the extinction in $V$, the
values of the S/N at 3 different wavelengths, roughly corresponding to
the middle of the range from the blue, green, and red arm of the STIS
spectrograph. There are 35 stars that SIMBAD recognises as spectroscopic binaries,
they are marked with a star (*). Finally, in this table we give the
references for the stars with parameters adopted from the literature.

\onltab{1}{
  \longtab{1}{
    \setlength{\tabcolsep}{4.5pt}
    \begin{longtable}{lrrrrrrrrrrrrc}
      \caption{\label{table:params} List of 367 stars in the NGSL. The columns
        are as follows: name; adopted atmospheric parameters and their
      errors (col 2 to 7); residual shifts in \kms; velocity
      dispersions in \kms, S/N at 3 different regions (col 10 - 12);
      extinction in $V$; reference for the atmospheric parameters. If
    not specified the measurements are from this paper.}\\
      \hline\hline
      Name & $T_{\rm eff}$ & error & $log(g)$ & error &{\rm  [Fe/H]}& error
      & cz & $\sigma$ &  \multicolumn{3}{c}{S/N @} \\
           & \multicolumn{2}{c}{(K)} &
           \multicolumn{2}{c}{($cm\,s^{-2}$)} &
           \multicolumn{2}{c}{(dex)} & \kms & \kms & 2800\,\AA&
           4000\,\AA\ & 8000\,\AA & A$_V$ & Reference \\
      \hline      
      \endfirsthead
      \caption{continued.}\\
      \hline\hline
      Name & $T_{\rm eff}$ & error & $log(g)$ & error &{\rm  [Fe/H]}& error
      & cz & $\sigma$ &  \multicolumn{3}{c}{S/N @} \\
           & \multicolumn{2}{c}{(K)} &
           \multicolumn{2}{c}{($cm\,s^{-2}$)} &
           \multicolumn{2}{c}{(dex)} & \kms & \kms & 2800\,\AA& 4000\,\AA\ & 8000\,\AA & A$_V$ & Reference \\
     \hline
      \endhead
      \hline
      \endfoot
           bd+092860 &  5298&    28&2.98&0.17&-1.99&0.06& -21& 129&  40&  94&  55& 0.06&                              \\
     bd+112998 &  5480&    30&3.00&0.24&-1.12&0.08&  -3& 182&  99& 251& 349& 0.11&                              \\
     bd+174708 &  6198&    16&4.18&0.06&-1.56&0.05& -20& 158& 193& 207& 101& 0.11&                              \\
     bd+292091 &  5615&    17&3.68&0.08&-2.01&0.04&  -5& 145& 116& 139&  72& 0.06&                              \\
     bd+371458 &  5365&    21&3.29&0.11&-2.01&0.05&  -6& 144& 160& 243& 141& 0.07&                              \\
     bd+381670 &  5535&    22&4.11&0.10&-0.70&0.06& -21& 162&  88& 219& 290& 0.07&                              \\
     bd+413306 &  5052&    38&4.41&0.14&-0.53&0.08& -40& 153&  44& 241& 393&-0.00&                              \\
     bd+413931 &  5530&    23&4.08&0.10&-1.48&0.06&  31& 145&  81& 129&  75& 0.12&                              \\
     bd+423607 &  5659&    19&3.69&0.09&-2.11&0.04&  17& 141& 127& 146&  73&-0.01&                              \\
     bd+442051 &  3664&    11&4.70&0.05&-0.83&0.04&  41& 178&   3& 127& 286& 0.20&                              \\
     bd+511696 &  5746&    16&4.39&0.06&-1.25&0.05&   1& 157& 103& 158&  89& 0.08&                              \\
     bd+592723 &  6035&    15&4.00&0.06&-1.94&0.04&   2& 161& 131& 129&  65& 0.18&                              \\
     bd+660268 &  5240&    29&3.45&0.17&-1.81&0.06&  17& 144&  76& 147&  86&-0.12&                              \\
     bd+720094 &  6174&    14&4.06&0.06&-1.76&0.04&  26& 170& 170& 169&  83& 0.11&                              \\
     bd+75d325 & 30086&  1531&3.59&0.29&+0.22&0.11&-111& 592& 201& 173& 202& 0.16&                              \\
     bd-122669*&  6892&    12&4.16&0.02&-1.47&0.03& -17& 192& 176& 155&  67& 0.14&                             0\\
     cd-259286 &  6336&    16&4.11&0.03&-1.22&0.03& -31& 139&  67& 117& 132& 0.29&                             0\\
    cd-3018140 &  6151&    17&4.01&0.07&-1.81&0.05&  -4& 154& 162& 167&  78& 0.02&                              \\
     cd-621346 &  5296&    42&2.86&0.30&-1.44&0.10&   7& 147&  69& 173& 246& 0.22&                              \\
     cd-691618 & 29000&     -&3.70&   -&-0.30&   -&  -1& 191& 820& 401& 206& 0.27&                           0,5\\
      g019-013 &  4083&    54&4.54&0.21&-0.58&0.19& -28& 147&  20& 272& 335&-0.32&                              \\
      g021-024 &  3995&    19&4.49&0.07&-0.44&0.07& -30& 147&  11& 144& 213& 0.06&                              \\
      g029-023 &  6143&    19&4.09&0.07&-1.74&0.05&  32& 154& 134& 144&  73& 0.21&                              \\
       g114-26*&  5966&    12&4.14&0.05&-1.59&0.04&   0& 157& 175& 188&  98& 0.06&                              \\
       g115-58 &  6117&    26&4.08&0.10&-1.65&0.07&   2& 148&  47&  56&  26& 0.15&                              \\
        g12-21 &  6021&    16&4.19&0.06&-1.43&0.05&  -7& 159& 120& 144&  74& 0.10&                              \\
        g13-35 &  6015&    18&3.99&0.07&-1.84&0.05& -20& 170& 186& 189&  96& 0.05&                              \\
       g169-28 &  5849&    18&4.21&0.07&-1.32&0.05&   7& 143&  54&  85&  44& 0.04&                              \\
        g17-25*&  5271&    23&4.44&0.09&-1.03&0.06&   9& 141&  49& 160& 111& 0.13&                              \\
        g18-39 &  6091&    17&4.18&0.06&-1.47&0.05&  23& 161& 115& 132&  69& 0.10&                              \\
        g18-54*&  6044&    21&4.23&0.08&-1.51&0.06&  12& 160&  94& 111&  59& 0.20&                              \\
       g180-24 &  6042&    13&4.13&0.05&-1.40&0.04&  12& 158& 150& 169&  87& 0.03&                              \\
       g187-40 &  5831&    18&4.20&0.07&-1.49&0.05&  24& 143&  89& 121&  64& 0.07&                              \\
       g188-22 &  6038&    14&4.16&0.06&-1.35&0.04&   5& 151& 128& 154&  79& 0.11&                              \\
       g188-30 &  5382&    24&4.11&0.11&-1.37&0.06&   9& 146&  32&  82&  48&-0.06&                              \\
       g192-43 &  6109&    18&4.05&0.07&-1.69&0.05& -22& 151& 128& 141&  68& 0.09&                              \\
       g194-22 &  5989&    20&4.07&0.08&-1.69&0.05&  -5& 146& 165& 183&  89&-0.06&                              \\
       g196-48*&  5767&    22&3.90&0.10&-1.75&0.06&  -5& 154& 101& 133&  72& 0.11&                              \\
        g20-15 &  6035&    20&4.12&0.08&-1.66&0.06&  -1& 153&  80& 112&  67& 0.57&                              \\
       g202-65*&  6656&    20&4.25&0.08&-1.37&0.07&  29& 171&  83&  84&  33&-0.10&                              \\
       g231-52 &  5414&    25&3.86&0.13&-1.67&0.06&  38& 159&  72& 122&  72& 0.01&                              \\
       g234-28*&  6066&    17&4.13&0.07&-1.58&0.05&  -6& 159&  81&  95&  47& 0.14&                              \\
         g24-3 &  5962&    21&4.06&0.09&-1.78&0.06&  19& 157& 110& 127&  67& 0.14&                              \\
       g243-62 &  4902&    17&4.68&0.07&-1.13&0.05& -21& 180&   5&  65& 122&-0.05&                              \\
       g260-36 &  4962&    30&4.45&0.10&+0.15&0.04& -18& 153&   7&  70& 134& 0.04&                              \\
       g262-14*&  5177&    24&4.44&0.09&-0.68&0.06&   2& 147&  10&  73& 119& 0.14&                              \\
        g63-26 &  5996&    28&3.92&0.12&-1.84&0.07& -11& 153&  46&  55&  25& 0.15&                              \\
        g88-27 &  6071&    17&4.12&0.07&-1.65&0.05&  -3& 162& 104& 115&  58& 0.07&                              \\
         gj825 &  3796&    18&4.55&0.08&-0.62&0.07&  98& 148&  27& 234& 573& 0.10&                              \\
         gl109 &  3462&     5&4.87&0.01&-0.20&0.20& 102& 142&   0&  53& 144& 0.11&                         0,7,9\\
         gl15b &  3630&    12&4.71&0.02&-0.88&0.03&  55& 165&   6& 191& 390& 0.16&                             0\\
      hd000319 &  8589&   339&4.45&0.29&-0.54&0.29&   5& 178& 790& 581& 535&-0.01&                              \\
      hd000358 & 12938&   366&4.19&0.57&+0.42&0.19&   0& 236& 904& 572& 470& 0.05&                              \\
      hd000886 & 20700&   822&3.81&0.33&-0.02&0.09&  -1& 160&1120& 639& 442& 0.10&                              \\
      hd001461 &  5588&    64&4.03&0.26&+0.08&0.10&  -1& 164& 211& 505& 624& 0.05&                              \\
      hd002665 &  5100&    45&2.62&0.30&-1.88&0.09&  37& 141& 141& 375& 278& 0.35&                              \\
      hd002857 &  7607&    12&3.78&0.09&-1.35&0.06&   3& 151& 127& 165&  58&-0.12&                              \\
      hd003360 & 20703&   984&3.83&0.41&-0.01&0.11&   0& 160&1455& 606& 452& 0.13&                              \\
      hd003712 &  4778&    63&2.06&0.33&+0.15&0.11& -16& 153&  97& 205& 412& 0.11&                              \\
      hd004128 &  4848&    62&2.25&0.36&+0.05&0.12&  -8& 149& 224& 276& 512& 0.03&                              \\
      hd004727*& 14851&   396&4.12&0.43&+0.14&0.17&   7& 210&1041& 639& 463& 0.17&                              \\
      hd004813 &  6167&    28&4.17&0.12&-0.18&0.06&   1& 144& 342& 467& 494& 0.12&                              \\
      hd005256 &  5170&    36&3.59&0.22&-0.65&0.07&   3& 150&  68& 271& 410& 0.09&                              \\
      hd005395 &  4888&    58&2.61&0.39&-0.42&0.12&  -4& 159& 124& 423& 657& 0.23&                              \\
      hd005544 &  4655&    54&2.26&0.35&-0.04&0.10& -11& 178&  19& 335& 629& 0.24&                              \\
      hd005916 &  4977&    57&2.69&0.42&-0.78&0.13&   1& 160&  97& 450& 642& 0.23&                              \\
      hd006229 &  5174&    32&2.58&0.24&-1.18&0.08&  10& 160&  83& 292& 446& 0.06&                              \\
      hd006734 &  4934&    46&3.18&0.32&-0.58&0.09& -13& 145& 111& 384& 498& 0.10&                              \\
      hd006755 &  5181&    33&2.84&0.24&-1.52&0.07&  32& 144& 162& 386& 266& 0.17&                              \\
      hd008491 &  4790&    50&2.60&0.31&+0.10&0.09& -14& 155& 120& 314& 535& 0.20&                              \\
      hd008724 &  4976&    39&2.43&0.30&-1.43&0.08&   7& 140&  36& 242& 237& 0.75&                              \\
      hd008890 &  6206&   110&1.64&0.47&+0.19&0.20&  -5& 142& 249& 383& 475& 0.12&                              \\
      hd009051 &  5036&    26&2.55&0.20&-1.51&0.06&  11& 143&  58& 209& 158& 0.23&                              \\
      hd010380 &  4191&    41&1.88&0.40&-0.14&0.10& -38& 141&  44& 244& 532& 0.24&                              \\
      hd010780 &  5400&    76&4.54&0.19&+0.12&0.11&  14& 151& 230& 462& 631& 0.17&                              \\
      hd012533*&  4402&    44&1.84&0.28&+0.12&0.08& -31& 138& 103& 184& 576& 0.55&                              \\
      hd013520*&  4054&    42&1.70&0.44&-0.13&0.12& -44& 156&  39& 274& 732& 0.46&                              \\
      hd015089 &  8796&   765&4.29&0.34&-0.94&0.63&  32& 128& 701& 523& 438& 0.01&                             0\\
      hd016031 &  6104&    14&4.07&0.06&-1.67&0.04& -14& 164& 177& 180&  91& 0.02&                              \\
      hd017072 &  5352&    53&2.83&0.40&-1.15&0.14&  -1& 151& 306& 526& 632& 0.01&                              \\
      hd017081*& 13893&   256&4.02&0.41&+0.21&0.13&  -3& 174&1101& 606& 474& 0.15&                              \\
      hd017361 &  4700&    44&2.67&0.30&+0.12&0.08& -17& 154& 103& 298& 519& 0.24&                              \\
      hd017925 &  5286&    81&4.69&0.19&+0.21&0.11&  -5& 153& 163& 366& 522& 0.29&                              \\
      hd018078 &  9791&   289&3.43&0.36&+1.00&0.07&   0& 159& 226& 456& 396& 0.43&                             0\\
      hd018769 &  8424&   206&4.34&0.12&-0.02&0.15&  34& 128& 673& 564& 516& 0.05&                             0\\
      hd018907 &  5059&    76&3.56&0.48&-0.72&0.15& -11& 150& 165& 528& 698& 0.08&                              \\
      hd019019 &  6033&    32&4.26&0.12&-0.18&0.07&   5& 150& 312& 477& 346& 0.15&                              \\
      hd019308 &  5697&    50&4.03&0.20&+0.06&0.08&   6& 149& 154& 453& 289& 0.13&                              \\
      hd019445 &  5805&    18&3.79&0.08&-2.04&0.04&  20& 160& 391& 400& 208& 0.08&                              \\
      hd019656 &  4717&    62&2.35&0.38&+0.05&0.11&  -9& 153& 102& 300& 520& 0.30&                              \\
      hd019787 &  4838&    54&2.61&0.34&+0.13&0.09&  -6& 152& 151& 318& 505& 0.17&                              \\
      hd020039*&  5206&    37&3.66&0.21&-0.77&0.08&  -2& 150&  65& 253& 387& 0.13&                              \\
      hd020630 &  5694&    58&4.39&0.18&+0.04&0.10&   5& 146& 290& 498& 614& 0.13&                              \\
      hd021742 &  5238&    73&4.21&0.25&+0.32&0.09&   2& 150&  63& 358& 588& 0.08&                              \\
      hd022049 &  5146&    74&4.68&0.19&+0.05&0.11&   2& 148& 206& 354& 493& 0.23&                              \\
      hd022484 &  5872&    49&3.97&0.23&-0.23&0.11&  -6& 168& 405& 542& 608& 0.04&                              \\
      hd023439*&  5198&    41&4.38&0.17&-0.90&0.10&  -5& 162&  77& 358& 522& 0.08&                              \\
      hd025329 &  5020&    40&4.42&0.19&-1.36&0.10&  11& 140&  54& 244& 199& 0.22&                              \\
      hd025893 &  5472&    66&4.62&0.16&+0.26&0.09&   0& 145& 114& 378& 335& 0.27&                              \\
      hd025975 &  4921&    67&3.29&0.40&+0.06&0.10&  -8& 145&  92& 350& 530& 0.15&                              \\
      hd026297 &  4827&    62&2.01&0.44&-1.51&0.11&  -9& 137&  30& 328& 355& 0.84&                              \\
      hd026630*&  5643&    84&1.54&0.44&+0.10&0.18&   0& 142& 391& 341& 564& 0.84&                              \\
      hd027295 & 11013&   337&4.06&0.50&+0.00&0.18&  -2& 212&1757& 630& 488& 0.07&                              \\
      hd028946 &  5338&    51&4.47&0.15&-0.08&0.09&  -4& 142&  82& 333& 240& 0.17&                              \\
      hd028978 &  8622&   723&3.73&1.39&-0.75&0.52& -20& 131& 666& 549& 483& 0.14&                              \\
      hd029391 &  7414&    31&4.09&0.13&-0.02&0.08&  -6& 161& 619& 614& 574& 0.09&                              \\
      hd029574 &  4712&    47&1.69&0.30&-1.52&0.10& -14& 140&   8& 175& 274& 1.46&                              \\
      hd030614 & 32902&  1917&3.31&0.23&+0.28&0.13&   1& 178& 590& 545& 533& 1.19&                              \\
      hd030834 &  4256&    40&1.67&0.33&-0.24&0.09& -36& 147&  35& 241& 575& 0.49&                              \\
      hd031219 &  6073&    49&4.12&0.18&+0.17&0.08&   6& 139& 146& 407& 476& 0.14&                              \\
      hd031421 &  4541&    37&2.43&0.28&-0.19&0.08& -39& 151& 105& 268& 516& 0.26&                              \\
      hd033793 &  3722&    28&4.71&0.05&-0.84&0.06&  40& 195&   3& 132& 645& 0.36&                             0\\
      hd034078 & 31057&  1696&3.54&0.25&-0.02&0.14&   6& 155&2571& 678& 572& 2.11&                              \\
      hd034797 & 12273&   355&4.23&0.50&+0.45&0.18&  -9& 232& 910& 624& 490&-0.01&                              \\
      hd034816 & 26885&  1880&3.60&0.29&-0.15&0.15&  -7& 188&1754& 674& 452& 0.29&                              \\
      hd036702 &  4768&    49&1.93&0.35&-1.78&0.09&   0& 135&  14& 205& 249& 1.14&                              \\
      hd036960 & 27000&     -&4.10&   -&-0.13&   -&  -9& 176&1338& 664& 466& 0.22&                         0,5,6\\
      hd037202 & 21132&  2024&3.24&0.41&+0.05&0.18& -55& 232&1486& 601& 453& 0.00&                              \\
      hd037216 &  5466&    54&4.56&0.14&+0.05&0.09&  10& 141&  96& 358& 246& 0.17&                              \\
      hd037763 &  4555&    72&3.17&0.49&+0.24&0.10& -16& 156&  72& 345& 687& 0.04&                              \\
      hd037828 &  4726&    76&2.04&0.60&-1.19&0.16& -14& 168&  38& 380& 671& 0.62&                              \\
      hd038237 &  8279&   471&4.32&0.63&-0.15&0.46& -11& 179& 667& 615& 529& 0.09&                              \\
      hd038510 &  5846&    29&3.96&0.16&-0.87&0.09& -21& 164& 220& 412& 488& 0.07&                              \\
      hd039587*&  5913&    41&4.34&0.14&-0.10&0.08&   8& 144& 298& 444& 507& 0.15&                              \\
      hd039833 &  5871&    49&4.39&0.15&+0.11&0.08&  -3& 147& 142& 422& 256& 0.15&                              \\
      hd040573 &  8795&   367&3.60&0.54&-1.39&0.27&   1& 182&1027& 612& 492&-0.03&                              \\
      hd041357*&  7838&    52&3.88&0.08&+0.35&0.05&   5& 180& 580& 549& 541& 0.19&                             0\\
      hd041661 &  6484&    40&3.98&0.18&-0.03&0.08&   0& 165& 287& 488& 257& 0.14&                              \\
      hd041667 &  4838&    84&2.31&0.65&-1.09&0.18&   1& 157&  27& 277& 536& 0.28&                              \\
      hd043042 &  6543&    31&4.16&0.12&-0.02&0.07&  -9& 148& 377& 476& 479& 0.15&                              \\
      hd044007 &  5085&    36&2.75&0.27&-1.50&0.08&  -2& 142&  86& 315& 246& 0.36&                              \\
      hd045282 &  5289&    33&3.16&0.24&-1.46&0.08& -17& 145& 174& 356& 229& 0.10&                              \\
      hd046703 &  6113&    28&4.02&0.13&-1.30&0.09&   6& 141&  77& 226& 134& 0.30&                              \\
      hd047839 & 32130&  1409&3.47&0.20&+0.13&0.10&   6& 227&1568& 718& 451& 0.51&                              \\
      hd048279 & 31593&  1806&3.51&0.23&+0.14&0.13&  -7& 176&1440& 626& 449& 1.63&                              \\
      hd050420 &  7265&    29&3.79&0.17&-0.00&0.08&  -1& 148& 475& 538& 585& 0.15&                              \\
      hd052089 & 22205&  2741&3.35&0.45&+0.01&0.18&  -2& 182& 989& 831& 406&-0.00&                              \\
      hd052973 &  5701&   102&1.32&0.49&+0.12&0.20&   8& 155& 238& 473& 668& 0.48&                              \\
      hd055057 &  7234&    33&3.87&0.18&+0.13&0.08&  -9& 162& 553& 566& 578& 0.04&                              \\
      hd055496 &  4935&    85&2.33&0.65&-1.44&0.16&  -8& 141&  51& 243& 207& 0.36&                              \\
      hd057060 & 32508&  1928&3.39&0.26&+0.24&0.14&  58& 247&2656& 577& 399& 0.72&                              \\
      hd057061 & 32514&   949&3.37&0.11&+0.18&0.07&   9& 222&2195& 601& 409& 0.63&                              \\
      hd057727 &  4966&    48&2.82&0.31&-0.17&0.09& -11& 148& 172& 367& 526& 0.13&                              \\
      hd058343 & 17497&  1268&3.29&0.55&+0.07&0.20& -17& 223& 708& 630& 562& 0.55&                              \\
      hd058551 &  6246&    23&4.21&0.11&-0.50&0.06&  -4& 176& 446& 575& 613& 0.09&                              \\
      hd059612 &  8306&   686&1.60&0.37&-0.20&0.46& -12& 157& 520& 556& 566& 0.16&                              \\
      hd060319 &  5907&    17&4.03&0.09&-0.82&0.05&  10& 166& 169& 297& 348& 0.08&                              \\
      hd061064 &  6568&    81&3.90&0.37&+0.06&0.16&  -9& 145& 548& 581& 602& 0.19&                              \\
      hd061603 &  3944&    38&1.53&0.47&+0.18&0.14& -16& 155&  21& 240& 692& 0.39&                              \\
      hd062412 &  4913&    56&2.67&0.35&+0.04&0.10& -17& 162& 135& 409& 644& 0.17&                              \\
      hd063077 &  5790&    36&4.00&0.19&-0.79&0.10& -17& 160& 424& 554& 608& 0.04&                              \\
      hd063700 &  5100&   143&0.90&0.51&+0.14&0.24& -31& 149& 215& 324& 652& 1.01&                              \\
      hd063791 &  5015&    37&2.57&0.28&-1.47&0.08&   5& 145&  64& 314& 270& 0.49&                              \\
      hd064412 &  5688&    25&4.05&0.12&-0.73&0.07& -11& 161& 136& 307& 382& 0.06&                              \\
      hd065228 &  5992&   118&1.43&0.52&+0.10&0.22&  -8& 152& 382& 507& 667& 0.27&                              \\
      hd065354 &  4146&    45&1.34&0.33&+0.07&0.11& -15& 153&  15& 242& 709& 0.79&                              \\
      hd065714 &  4908&   105&2.21&0.57&+0.15&0.19& -39& 156& 104& 398& 619& 0.10&                              \\
      hd067390 &  7142&    15&3.96&0.08&-0.00&0.04& -10& 157& 155& 237& 234& 0.18&                              \\
      hd068988 &  5755&    47&4.00&0.20&+0.22&0.07&   4& 142& 112& 371& 484& 0.08&                              \\
      hd071160 &  4097&    30&1.87&0.34&+0.07&0.08& -23& 153&   5& 204& 718& 0.44&                              \\
      hd072184*&  4643&    71&2.84&0.47&+0.23&0.11& -35& 158&  58& 358& 665& 0.11&                              \\
      hd072324 &  4858&    86&2.32&0.50&+0.05&0.16& -32& 163&  75& 384& 597& 0.15&                              \\
      hd072505 &  4596&    87&2.81&0.59&+0.27&0.13& -32& 159&  36& 343& 660& 0.24&                              \\
      hd072968 &  9645&  1534&3.60&4.13&+0.83&0.43&  39& 228& 639& 574& 510&-0.25&                              \\
      hd073710 &  4906&    75&2.54&0.41&+0.23&0.12&  -5& 161&  80& 392& 628& 0.17&                              \\
      hd074088 &  4015&    45&1.69&0.46&-0.26&0.13& -31& 155&   7& 233& 884& 0.88&                              \\
      hd074721 &  8475&    96&3.35&0.25&-1.47&0.11&   1& 153& 420& 417& 295& 0.05&                              \\
      hd076291 &  4609&    60&2.81&0.44&-0.03&0.11& -22& 155&  61& 360& 659& 0.21&                              \\
      hd076932 &  5894&    30&4.07&0.15&-0.90&0.09&  -8& 163& 444& 567& 625& 0.08&                              \\
      hd078316*& 12279&   613&2.94&0.57&+0.22&0.18& -11& 188& 883& 622& 481& 0.03&                              \\
      hd078362*&  7343&   100&3.86&0.43&+0.57&0.15&   5& 141& 461& 539& 568& 0.19&                              \\
      hd078479 &  4580&    95&2.87&0.64&+0.33&0.13& -13& 160&  19& 329& 665& 0.28&                              \\
      hd079158 & 12737&   717&3.09&0.68&+0.49&0.18& -19& 160& 872& 625& 480&-0.07&                              \\
      hd079349 &  3884&    19&1.79&0.26&+0.04&0.09& -12& 153&   2& 147& 681& 0.34&                              \\
      hd079469 &  8691&   234&3.23&0.34&-1.48&0.18&   4& 169& 913& 572& 481&-0.12&                              \\
      hd080607 &  5389&    45&3.99&0.18&+0.35&0.06&   2& 152&  43& 221& 341& 0.04&                              \\
      hd081797 &  4186&    39&1.73&0.35&+0.07&0.09& -20& 142&  53& 180& 650& 0.38&                              \\
      hd082395 &  4823&    74&2.79&0.47&+0.04&0.13& -16& 146& 107& 327& 544& 0.32&                              \\
      hd082734 &  4935&    81&2.50&0.43&+0.26&0.13&  -7& 164& 154& 395& 637& 0.16&                              \\
      hd083212 &  4763&    58&1.79&0.39&-1.39&0.11& -10& 138&  21& 228& 232& 0.44&                              \\
      hd085380 &  5957&    48&3.98&0.22&-0.06&0.09&   3& 145& 289& 466& 633& 0.11&                              \\
      hd086322 &  4804&    46&2.53&0.30&-0.05&0.09&  -4& 151&  44& 336& 424& 0.22&                              \\
      hd086986 &  8031&    84&3.57&0.45&-1.47&0.19&   3& 160& 482& 547& 428& 0.13&                              \\
      hd087140 &  5145&    20&2.75&0.14&-1.73&0.04&  15& 145&  97& 218& 151& 0.09&                              \\
      hd087737 & 11522&   640&2.11&0.37&+0.19&0.16& -16& 142&1190& 628& 486& 0.03&                              \\
      hd090862 &  4129&    36&1.70&0.36&-0.39&0.09& -55& 154&   3& 157& 599& 0.52&                              \\
      hd091316 & 21576&   668&3.01&0.05&-0.06&0.06&   4& 181& 959& 621& 476& 0.07&                             0\\
      hd093329 &  8127&    84&3.45&0.41&-1.45&0.17& -22& 160& 351& 390& 292& 0.05&                              \\
      hd093813 &  4456&    41&2.36&0.33&-0.10&0.09& -25& 144& 114& 247& 527& 0.35&                              \\
      hd094028 &  5982&    17&4.09&0.07&-1.54&0.05&  -5& 167& 316& 365& 191& 0.06&                              \\
      hd095241*&  5778&    37&3.78&0.21&-0.47&0.09&   0& 151& 333& 459& 509& 0.09&                              \\
      hd095418 &  8734&   542&3.68&1.12&-0.76&0.32&  -1& 141& 723& 441& 425&-0.03&                              \\
      hd095735 &  3574&    12&4.73&0.07&-0.93&0.05&   8& 157&   9& 237& 535&-0.02&                              \\
      hd095849 &  4526&    48&2.38&0.32&+0.21&0.08&  -5& 152&  33& 329& 661& 0.19&                              \\
      hd096446 & 20086&   530&3.59&0.08&+0.06&0.04&  -7& 202& 915& 643& 486& 0.23&                             0\\
      hd097633 &  8790&   351&3.59&0.89&-0.64&0.18& -11& 151& 805& 582& 481&-0.04&                              \\
      hd099648 &  4970&    75&2.25&0.43&-0.01&0.15& -10& 144& 153& 403& 618& 0.28&                              \\
      hd101013*&  5043&   332&2.93&2.06&+0.12&0.54&   2& 172& 126& 384& 587& 0.71&                              \\
      hd101107 &  7036&    16&4.09&0.08&-0.02&0.04&  -7& 208& 781& 538& 562& 0.10&                              \\
      hd102212 &  3738&     6&1.55&0.10&-0.41&0.05&  -9& 150&  36& 218& 539& 0.12&                            12\\
      hd102780 &  3835&    22&1.64&0.31&-0.20&0.12& -24& 156&   3& 162& 751& 0.53&                              \\
      hd103036 &  4688&    80&1.64&0.53&-1.40&0.17&   6& 145&  21& 225& 273& 0.85&                              \\
      hd105452 &  7049&    19&4.11&0.10&-0.21&0.06&  -9& 181& 588& 578& 571& 0.09&                              \\
      hd105546 &  5242&    31&2.73&0.22&-1.46&0.07&  11& 138& 105& 264& 173& 0.05&                              \\
      hd105740 &  4771&    34&2.77&0.26&-0.69&0.07& -21& 180&  26& 263& 520& 0.18&                              \\
      hd106304 &  8675&   205&2.85&0.18&-1.63&0.11&  -6& 160& 362& 358& 246& 0.04&                              \\
      hd106516*&  6236&    26&4.20&0.12&-0.70&0.08&   2& 167& 500& 581& 607& 0.07&                              \\
      hd107582*&  5540&    36&4.13&0.16&-0.72&0.09&  10& 152& 157& 393& 506& 0.06&                              \\
      hd108945 &  8906&  1668&4.19&1.54&-1.48&1.36&  37& 128& 613& 568& 498& 0.03&                              \\
      hd109387 & 16906&  1478&3.19&0.62&-0.12&0.31& -47& 321&1965& 526& 417& 0.29&                              \\
      hd109995 &  8427&   174&3.41&0.45&-1.52&0.18&  22& 162& 661& 602& 502& 0.09&                              \\
      hd110073 & 13000&     -&3.90&   -&-0.40&0.30& -10& 175& 975& 620& 463& 0.27&                         0,3,5\\
      hd110885 &  5528&    29&3.09&0.22&-1.21&0.08&  16& 147& 100& 214& 130& 0.11&                              \\
      hd111464 &  4314&    43&2.17&0.40&-0.03&0.09& -30& 162&  13& 284& 713& 0.64&                              \\
      hd111515 &  5373&    48&4.23&0.19&-0.59&0.11&   1& 154& 111& 379& 535& 0.07&                              \\
      hd111721 &  5120&    37&2.90&0.28&-1.27&0.08&   9& 141& 102& 333& 248& 0.22&                              \\
      hd111786 &  7549&    45&4.17&0.21&-1.06&0.21&   3& 190& 774& 600& 564& 0.09&                              \\
      hd112413 & 11658&   471&3.78&1.19&+0.68&0.17&  -1& 163&1036& 541& 452&-0.08&                              \\
      hd113002 &  5152&    37&2.53&0.28&-1.08&0.09&   0& 153&  68& 271& 414& 0.04&                              \\
      hd113092 &  4319&    38&1.53&0.33&-0.70&0.09& -24& 152&  35& 318& 701& 0.24&                              \\
      hd114330 &  9671&   289&3.57&0.82&-0.24&0.12&  -5& 134& 641& 429& 373& 0.24&                              \\
      hd114710 &  5973&    49&4.23&0.18&-0.04&0.09&   2& 147& 378& 532& 593& 0.11&                              \\
      hd115617 &  5506&    57&4.30&0.18&-0.03&0.10&   0& 144& 228& 410& 500& 0.13&                              \\
      hd117880 &  9000&     -&3.01&0.03&-1.62&0.03& -14& 143& 344& 354& 247& 0.23&                        0,15,4\\
      hd118055 &  4717&    38&1.74&0.25&-1.57&0.07&   0& 138&   8& 148& 198& 1.13&                              \\
      hd119971 &  4233&    42&1.67&0.43&-0.61&0.10& -49& 144&  24& 309& 696& 0.28&                              \\
      hd121146 &  4454&    47&2.92&0.41&+0.03&0.09& -30& 158&  26& 280& 569& 0.20&                              \\
      hd122064 &  4490&    68&4.30&0.37&+0.09&0.11&  -1& 147&  57& 314& 527&-0.22&                              \\
      hd122956 &  4932&    61&2.33&0.47&-1.47&0.12& -11& 148&  61& 376& 376& 0.62&                              \\
      hd123657 &  3261&    43&0.59&0.38&-0.02&0.19& -15& 165&  39& 246& 987& 0.28&                            13\\
      hd124186 &  4458&    57&2.79&0.44&+0.31&0.08& -25& 154&  25& 315& 642& 0.21&                              \\
      hd124425*&  6355&    39&4.01&0.18&-0.10&0.09&  -1& 148& 521& 555& 578& 0.11&                              \\
      hd124547*&  4165&    48&1.73&0.45&-0.19&0.11& -47& 142& 222& 244& 574& 0.19&                              \\
      hd126327 &  3100&     0&1.98&6.27&-0.45&3.47&  22& 175&   7& 313& 424& 0.27&                              \\
      hd126511 &  5402&    38&4.16&0.13&+0.17&0.06&   1& 160&  66& 272& 193& 0.07&                              \\
      hd126614 &  5453&    59&3.87&0.25&+0.53&0.07&  -5& 158&  42& 245& 392& 0.05&                              \\
      hd126661 &  7809&    91&3.83&0.35&+0.36&0.13&  -1& 169& 613& 560& 547& 0.11&                              \\
      hd128000 &  3954&    33&1.75&0.41&+0.09&0.12& -21& 154&  21& 256& 744& 0.38&                              \\
      hd128279 &  5279&    22&3.07&0.13&-2.08&0.04&  16& 150& 219& 356& 229& 0.22&                              \\
      hd128801 &  8811&   283&2.55&0.16&-1.69&0.11&  16& 185& 492& 424& 269&-0.09&                              \\
      hd128987 &  5638&    69&4.63&0.16&+0.09&0.11&   0& 147& 132& 424& 306& 0.22&                              \\
      hd131873 &  4077&    28&1.70&0.29&-0.10&0.08& -15& 151&  86& 179& 672& 0.39&                              \\
      hd132345 &  4484&    67&2.58&0.47&+0.37&0.09& -23& 165&  31& 316& 634& 0.16&                              \\
      hd132475 &  5721&    20&3.79&0.10&-1.61&0.05& -13& 150& 220& 296& 170& 0.10&                              \\
      hd134113*&  5668&    24&3.85&0.14&-0.82&0.07&   4& 165& 171& 384& 465& 0.06&                              \\
      hd134439 &  5357&    35&4.68&0.10&-1.11&0.08&  38& 134&  51& 180& 121&-0.04&                              \\
      hd134440 &  5094&    36&4.70&0.12&-1.09&0.09&  30& 154&  31& 158& 128& 0.20&                              \\
      hd136726 &  4235&    37&2.00&0.35&+0.02&0.08& -43& 147&  34& 239& 551& 0.31&                              \\
      hd137759 &  4525&    46&2.52&0.34&+0.13&0.08& -20& 147& 120& 270& 388&-0.27&                              \\
      hd137909 &  8620&   499&3.96&1.18&+1.00&0.00&  -9& 154& 627& 495& 483& 0.23&                              \\
      hd138716 &  4767&    54&3.05&0.36&-0.01&0.09& -15& 150& 123& 312& 520& 0.15&                              \\
      hd138749 & 16150&   346&3.75&0.38&+0.13&0.11&   4& 220&1063& 605& 439& 0.23&                              \\
      hd140232 &  8381&   238&4.27&0.17&+0.26&0.15&   6& 147& 695& 572& 531& 0.18&                             0\\
      hd141795 &  8516&   184&4.48&0.09&-0.17&0.14&  41& 128& 753& 552& 518& 0.01&                             0\\
      hd141851 &  8524&   383&4.27&0.39&-0.47&0.30&   4& 222& 771& 572& 521& 0.10&                              \\
      hd142091 &  4769&    54&2.97&0.36&+0.07&0.09&  -4& 149& 135& 328& 516& 0.07&                              \\
      hd142703 &  7235&    26&4.12&0.15&-1.20&0.13&  -3& 185& 722& 590& 580& 0.06&                              \\
      hd142860 &  6275&    29&4.12&0.13&-0.26&0.07&   7& 164& 383& 490& 476& 0.11&                              \\
      hd142926 & 12831&   782&3.33&1.03&+0.27&0.24&  -8& 235& 806& 619& 484& 0.19&                              \\
      hd143107 &  4460&    44&2.20&0.34&-0.11&0.09& -16& 137&  75& 269& 508& 0.29&                              \\
      hd143459 & 10298&   277&3.85&0.50&-0.47&0.12&   2& 182& 940& 588& 514& 0.39&                              \\
      hd145328 &  4783&    59&2.96&0.40&-0.00&0.10&  -6& 157& 161& 394& 643& 0.14&                              \\
      hd146051 &  3783&    20&1.45&0.19&-0.03&0.06& -10& 146&  71& 230& 572& 0.28&                            13\\
      hd146233 &  5696&    57&4.20&0.21&-0.06&0.10&   1& 148& 250& 433& 486& 0.11&                              \\
      hd147394 & 14906&   332&4.06&0.39&+0.14&0.15&  -5& 201&1154& 636& 469& 0.12&                              \\
      hd147550 &  9830&   279&3.70&0.66&-0.38&0.11&  -1& 153& 960& 599& 515& 0.42&                              \\
      hd148293 &  4695&    62&2.37&0.38&+0.20&0.11&  -8& 158&  79& 361& 647& 0.14&                              \\
      hd148513 &  4147&    41&2.13&0.48&+0.21&0.09& -18& 152&  21& 263& 638& 0.50&                              \\
      hd149161 &  3951&    25&1.79&0.28&-0.18&0.09& -13& 153&  24& 214& 599& 0.36&                              \\
      hd149382 & 27535&  1985&3.92&0.76&-0.55&0.16& -36& 522&1491& 500& 218& 0.44&                              \\
      hd155763 & 14035&   332&3.57&0.44&+0.22&0.11&  -3& 176&1002& 605& 448& 0.19&                              \\
      hd156283 &  4274&    45&1.78&0.35&+0.10&0.09& -32& 151&  27& 259& 796& 0.50&                              \\
      hd157244 &  4479&    89&1.37&0.37&+0.23&0.14& -35& 153& 182& 270& 634& 0.86&                              \\
      hd159181 &  5325&    87&1.51&0.47&-0.02&0.20& -15& 159& 362& 431& 553& 0.44&                              \\
      hd160346*&  4808&    65&4.53&0.22&+0.03&0.10& -15& 154&  84& 336& 512& 0.05&                              \\
      hd160762 & 17789&   604&3.87&0.46&+0.00&0.13&   0& 169&1018& 635& 468& 0.19&                              \\
      hd160922*&  6595&    28&4.19&0.11&-0.03&0.06&  -1& 160& 494& 564& 595& 0.12&                              \\
      hd161770 &  5782&    20&3.95&0.09&-1.60&0.05&   5& 169&  98& 163& 108& 0.56&                              \\
      hd163346 &  6910&   133&4.02&0.55&+0.23&0.23&  90& 250& 418& 430& 387& 0.69&                              \\
      hd163641 & 11953&   263&4.06&0.44&+0.19&0.15&   2& 188&1283& 609& 519& 0.41&                              \\
      hd163810 &  5818&    15&4.35&0.06&-1.20&0.04& -11& 148& 102& 156&  86& 0.11&                              \\
      hd164058 &  3985&    32&1.69&0.38&+0.11&0.11& -18& 164& 133& 237& 664& 0.34&                              \\
      hd164257 &  9792&   691&3.70&2.11&+0.41&0.30& -10& 178& 720& 610& 526& 0.38&                              \\
      hd164353 & 17574&   662&2.96&0.23&+0.15&0.11&  -9& 155&3635& 608& 506& 0.48&                              \\
      hd164402*& 29405&   374&3.30&0.02&+0.02&0.03&  -8& 158&5697& 707& 522& 0.93&                             0\\
      hd164967 &  9001&   548&4.34&0.36&-1.32&0.43&  36& 128& 794& 616& 523& 0.14&                              \\
      hd165195 &  4766&    59&1.89&0.40&-1.98&0.09&   7& 134&  20& 303& 428& 1.54&                              \\
      hd165341*&  5365&    62&4.49&0.16&+0.19&0.09&  -2& 141& 185& 339& 462& 0.24&                              \\
      hd166208*&  4953&    75&2.19&0.46&-0.06&0.16& -13& 160& 292& 419& 595&-0.04&                              \\
      hd166229 &  4577&    67&2.82&0.47&+0.23&0.10& -34& 177&  61& 349& 644& 0.10&                              \\
      hd166283 &  8574&   616&4.55&0.53&-0.41&0.57&  29& 128& 459& 571& 473& 0.21&                              \\
      hd166991 &  8977&  1010&4.40&0.55&-1.39&0.79&  27& 147& 844& 606& 511& 0.02&                              \\
      hd167006 &  3535&    24&0.99&0.29&-0.08&0.10&  41& 148&  28& 188& 735& 0.38&                            13\\
      hd167105 &  8637&   143&3.25&0.24&-1.52&0.10&  15& 143& 325& 356& 235& 0.07&                              \\
      hd167278 &  6563&    18&4.14&0.08&-0.21&0.04&   1& 149& 245& 363& 173& 0.12&                              \\
      hd167946 &  9300&   502&3.74&0.87&-0.77&0.20&   0& 163&1024& 616& 494& 0.27&                              \\
      hd169191 &  4426&    41&2.22&0.33&-0.12&0.09& -26& 143&  44& 272& 554& 0.35&                              \\
      hd170737*&  5042&    44&3.25&0.31&-0.87&0.09&  11& 154&  69& 346& 564& 0.13&                              \\
      hd170756 &  5903&   106&3.79&0.61&-1.17&0.32&  23& 140& 107& 461& 656& 0.74&                              \\
      hd170973 & 12046&   925&3.56&1.74&+0.73&0.28&  -6& 140& 816& 587& 500& 0.18&                              \\
      hd172230 &  7772&   102&3.76&0.44&+0.55&0.14&   0& 188& 458& 587& 528& 0.12&                              \\
      hd172506 &  7078&    20&4.00&0.10&-0.15&0.06&  -5& 141& 304& 427& 181& 0.06&                              \\
      hd173158 &  5164&   121&0.87&0.43&+0.04&0.20&  -1& 154&  10& 206& 857& 2.28&                              \\
      hd173819 &  3650&   102&0.51&1.38&-0.11&0.88&   5& 179&  17& 212& 672&-0.79&                              \\
      hd174240 &  8879&   340&3.61&0.88&-0.63&0.16&  -7& 166& 809& 583& 514& 0.12&                              \\
      hd174959 & 14321&   348&4.05&0.48&+0.18&0.17&   3& 195&1054& 636& 485& 0.27&                              \\
      hd174966 &  7874&    57&4.09&0.16&+0.03&0.10&  -9& 152& 430& 575& 505& 0.32&                              \\
      hd175156 & 16361&   556&3.04&0.25&+0.17&0.10&   1& 139& 465& 523& 572& 1.01&                              \\
      hd175305 &  5118&    42&2.75&0.31&-1.43&0.09&  15& 139& 151& 426& 323& 0.05&                              \\
      hd175545 &  4526&    53&2.95&0.40&+0.12&0.09& -24& 161&  16& 337& 674& 0.26&                              \\
      hd175640 & 12067&   326&4.07&0.55&+0.22&0.18&   0& 176& 829& 608& 496& 0.29&                              \\
      hd175674 &  4421&   114&2.44&0.90&+0.21&0.20& -27& 161&  18& 294& 722& 0.57&                              \\
      hd175805 &  6273&    47&3.98&0.21&+0.15&0.08&   1& 144& 195& 445& 239& 0.19&                              \\
      hd175865 &  3181&    52&0.47&0.47&-0.29&0.33&  -9& 156&  74& 224& 800& 0.14&                            13\\
      hd176232 &  8659&   308&4.47&0.33&+0.55&0.17&   0& 153& 585& 552& 515& 0.17&                              \\
      hd176437 & 12715&   367&3.68&0.67&+0.17&0.15&   0& 128& 844& 534& 496& 0.38&                              \\
      hd181720 &  5659&    42&3.88&0.23&-0.65&0.10&   9& 152& 191& 481& 598& 0.06&                              \\
      hd183324 &  8939&   304&4.53&0.23&-1.30&0.50&  13& 312& 764& 550& 474&-0.26&                              \\
      hd183915 &  4091&   129&0.91&1.07&-1.17&0.42&  -1& 145&  23& 309& 735&-0.41&                              \\
      hd184266 &  5700&     -&2.00&   -&-1.65&   -&  35& 147& 239& 452& 270& 0.41&                    0,14,15,16\\
      hd185144 &  5283&    72&4.51&0.20&-0.11&0.12&   7& 151& 204& 384& 515& 0.17&                              \\
      hd185351 &  4921&    56&2.95&0.36&+0.01&0.09&  -2& 143& 163& 356& 519& 0.09&                              \\
      hd187111 &  4764&    59&1.93&0.42&-1.44&0.11&   0& 142&  18& 269& 339& 1.07&                              \\
      hd187879 & 20420&  1274&3.18&0.28&-0.02&0.14& -12& 197& 688& 608& 517& 0.40&                              \\
      hd188262 &  5749&   108&2.80&0.68&+0.15&0.20&   4& 149& 369& 375& 285& 0.62&                              \\
      hd190073 & 10900&   546&4.13&0.66&-0.02&0.26& 181& 273& 434& 610& 463& 0.60&                              \\
      hd190360 &  5427&    74&3.93&0.31&+0.20&0.10&  -4& 180& 282& 485& 610& 0.05&                              \\
      hd190404 &  4982&    55&4.49&0.20&-0.62&0.13& -10& 146&  83& 395& 332& 0.04&                              \\
      hd191026 &  5133&    69&3.77&0.33&+0.04&0.10&  -4& 167& 200& 370& 518& 0.19&                              \\
      hd191277 &  4462&    50&2.89&0.41&+0.16&0.08& -26& 156&  45& 328& 667& 0.12&                              \\
      hd193281 &  8623&   345&4.30&0.33&-0.68&0.28&  -4& 192& 755& 544& 489& 0.03&                              \\
      hd193495*&  5458&   125&2.55&0.65&+0.13&0.21&  10& 161& 868& 417& 669& 0.18&                              \\
      hd194093 &  6000&     -&0.85&0.10&+0.15&0.09&  11& 155& 354& 473& 442& 0.72&                  0,8,10,11,12\\
      hd194453 & 10342&   294&3.69&0.79&-0.06&0.14&  -6& 180& 838& 591& 504& 0.23&                              \\
      hd195434 &  4858&    22&4.40&0.09&-0.57&0.05& -46& 177&  40& 184& 348& 0.09&                              \\
      hd196218 &  6207&    24&4.11&0.11&-0.19&0.05&   0& 150& 279& 487& 271& 0.10&                              \\
      hd196426 & 12951&   187&4.10&0.27&+0.22&0.10&   0& 204& 785& 517& 413& 0.16&                              \\
      hd196662 & 15439&    90&3.90&0.06&+0.14&0.03&  -9& 179& 596& 474& 363& 0.29&                             0\\
      hd196725 &  4260&    61&1.18&0.32&+0.06&0.13& -26& 160&  38& 256& 744& 0.66&                              \\
      hd196892 &  6028&    22&4.17&0.10&-0.99&0.07&   6& 165& 257& 412& 469& 0.06&                              \\
      hd197177 &  4955&    74&2.07&0.40&+0.02&0.14&   0& 162& 120& 373& 619& 0.31&                              \\
      hd198809 &  5075&    43&2.54&0.28&-0.27&0.09&   5& 154& 264& 381& 514& 0.02&                              \\
      hd200081 &  5526&    71&3.25&0.43&+0.02&0.12&   9& 147& 159& 322& 241& 0.40&                              \\
      hd200905 &  3997&    39&0.92&0.26&+0.12&0.12& -42& 148& 271& 193& 600& 0.54&                              \\
      hd201091 &  4167&    56&4.54&0.28&-0.35&0.20&  20& 156&  75& 330& 695&-0.12&                              \\
      hd201377 &  8415&   542&4.32&0.62&-0.23&0.46&   0& 166& 731& 589& 518& 0.12&                              \\
      hd201601 &  8574&   406&4.47&0.51&+0.68&0.22&   9& 163& 759& 591& 511& 0.26&                              \\
      hd203638 &  4647&    92&2.81&0.59&+0.27&0.14& -23& 166&  55& 353& 654& 0.26&                              \\
      hd204041 &  8617&   295&4.49&0.25&-0.61&0.28&   6& 174& 828& 578& 545& 0.04&                              \\
      hd204155 &  5704&    28&3.89&0.16&-0.70&0.07&   5& 172& 151& 349& 430& 0.03&                              \\
      hd204543 &  4874&    50&1.99&0.34&-1.78&0.08&  15& 134&  40& 256& 224& 0.34&                              \\
      hd204867 &  5715&   100&1.22&0.46&+0.10&0.19&  -4& 144& 219& 385& 489& 0.38&                              \\
      hd205202 &  6496&    22&4.07&0.10&-0.51&0.06&   0& 170& 286& 456& 473& 0.07&                              \\
      hd205811 &  9069&   753&4.42&0.42&-1.30&0.61&  23& 146&1203& 683& 484&-0.23&                              \\
      hd206778 &  4240&    47&0.93&0.21&+0.08&0.10& -10& 145& 226& 142& 558& 0.67&                              \\
      hd210745 &  4337&    81&1.12&0.33&+0.17&0.14& -37& 157& 157& 246& 632& 0.89&                              \\
      hd210807 &  5023&    54&2.31&0.33&-0.16&0.11&   1& 143& 189& 350& 516& 0.21&                              \\
      hd212516 &  3709&    11&1.54&0.20&-0.24&0.09&  -7& 150&   1& 118& 677& 0.51&                              \\
      hd212593 & 13642&   251&2.42&0.06&+0.30&0.04&  -6& 158&1228& 628& 526& 0.50&                             0\\
      hd215665 &  4933&    78&2.25&0.42&+0.12&0.14&   7& 152& 145& 313& 510& 0.41&                              \\
      hd217107 &  5495&    63&3.99&0.25&+0.30&0.08&  -8& 165& 190& 480& 615& 0.03&                              \\
      hd217357 &  3894&    25&4.48&0.10&-0.47&0.10& -16& 149&  20& 245& 703& 0.13&                              \\
      hd221377 &  6399&    28&4.08&0.13&-0.66&0.08& -12& 176& 439& 588& 614& 0.09&                              \\
      hd222404*&  4734&    56&3.10&0.37&+0.13&0.08& -14& 159& 334& 356& 565& 0.08&                              \\
      hd224801 & 12704&   568&4.11&0.94&+0.69&0.21&  -4& 188&1191& 604& 490& 0.13&                              \\
      hd224926 & 14120&   317&4.06&0.51&+0.31&0.16&  -7& 192& 915& 626& 481& 0.15&                              \\
      hd232078 &  4295&    48&0.82&0.27&-1.08&0.11& -11& 146&   1& 106& 333& 2.21&                              \\
      hd284248 &  6098&    16&4.12&0.06&-1.60&0.05& -27& 157& 228& 235& 116& 0.05&                              \\
      hd345957 &  5883&    17&4.02&0.07&-1.45&0.05&  11& 146& 201& 257& 143& 0.08&                              \\
        hr0753 &  4529&    74&4.40&0.38&-0.21&0.15& -13& 149&  67& 300& 446&-0.45&                              \\
        hr8086 &  3894&    27&4.54&0.09&-0.56&0.09& -15& 148&  39& 239& 563& 0.03&                              \\
         lhs10 &  3167&    12&5.34&0.03&-0.42&0.10& -36& 128&   4&  14& 235& 2.26&                             0\\
        lhs482 &  3707&    17&4.83&0.07&-0.78&0.07& -53& 162&   1&  27& 154& 0.38&                              \\
       mmj6476 &  7648&    22&3.96&0.09&+0.28&0.04& -11& 156&  66& 106&  98& 0.27&                              \\
       mmj6490 &  8947&   211&4.36&0.15&-1.37&0.21&  14& 135& 130& 142& 104& 0.02&                              \\
        vbnvul &  7944&    44&1.85&0.01&-1.58&0.04&  52& 163&  32& 102&  70& 1.67&                             0\\
        vgkcom &  3254&    16&0.60&0.12&-1.93&   -& -10& 149&  11& 229& 926& 0.27&                           0,1\\
        viwcom &  3303&    28&0.61&0.46&+0.08&0.34& -18& 146&   6& 166& 516& 0.27&                              \\

      \hline
    \end{longtable}
    \tablefoot{ References: 0 - \cite{heap2009}, 1 - \cite{1}, 2 -
      \cite{2}, 3 - \cite{3}, 4 - \cite{4}, 5 -
      \cite{cayrel2001}, 6 - \cite{6}, 7 - \cite{7}, 8 -
      \cite{8}, 9 -\cite{9}, 10 -\cite{10}, 11 - \cite{11}, 12 -
      \cite{wu2011}, 13 - \cite{miles_params}, 14 - \cite{14}, 15
      -\cite{15}, 16 - \cite{16};  } 
  } 
} 

\section{Comparison with other libraries}
\label{app:comparison}

Here we plot the comparisons with the literature discussed in
Sect.\,\ref{sec:comp}. For each of the reference libraries we plot
the different spectroscopic classes in different panels. We omit the M
class because there are too few (14) cold stars in NGSL. We provide the
usual `our' \emph{vs.} `literature' value plots, but we also
investigate how the residuals of this comparison depend on the
different parameters.  

\begin{figure*}
\includegraphics[width=0.99\textwidth]{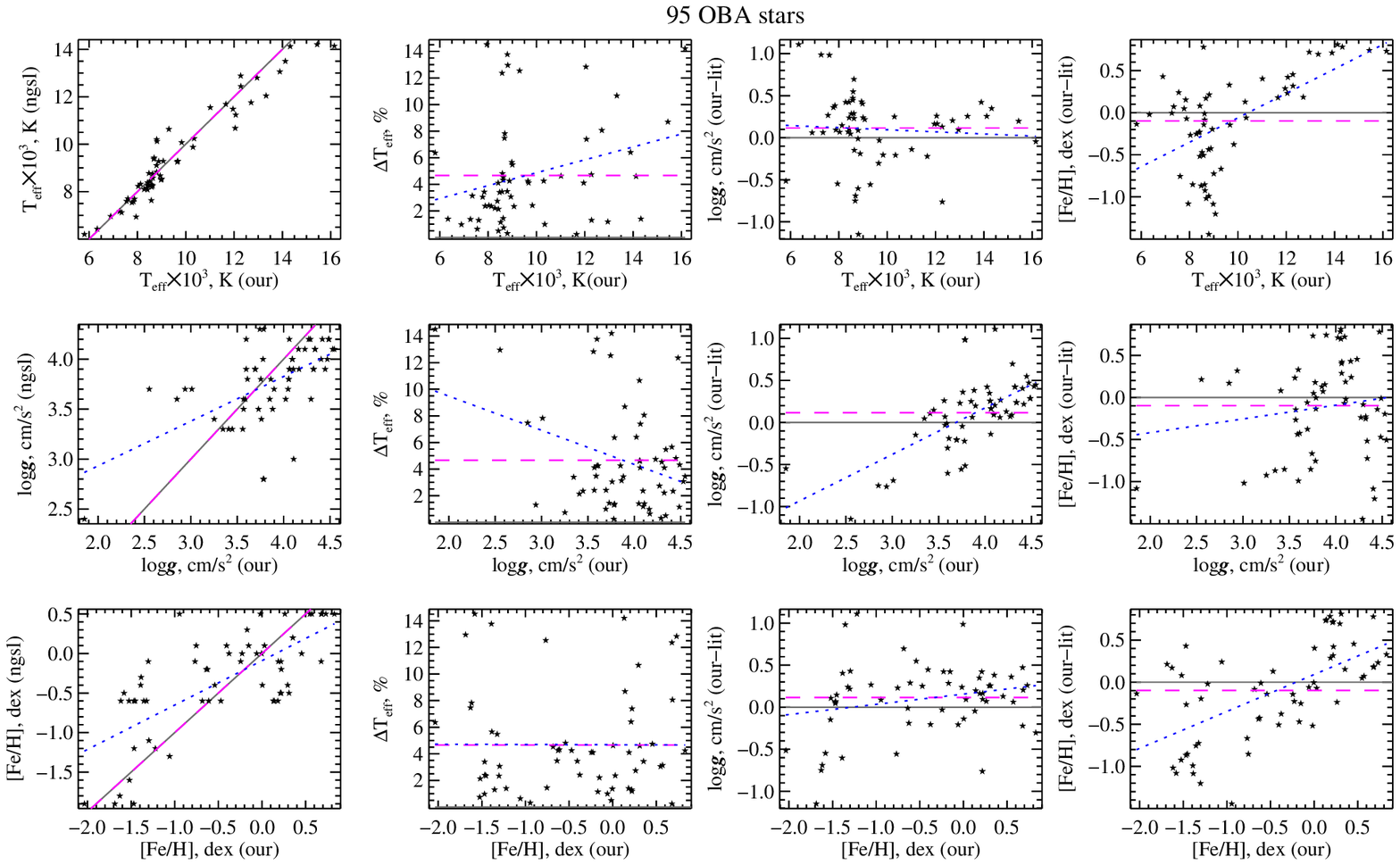}
\includegraphics[width=0.99\textwidth]{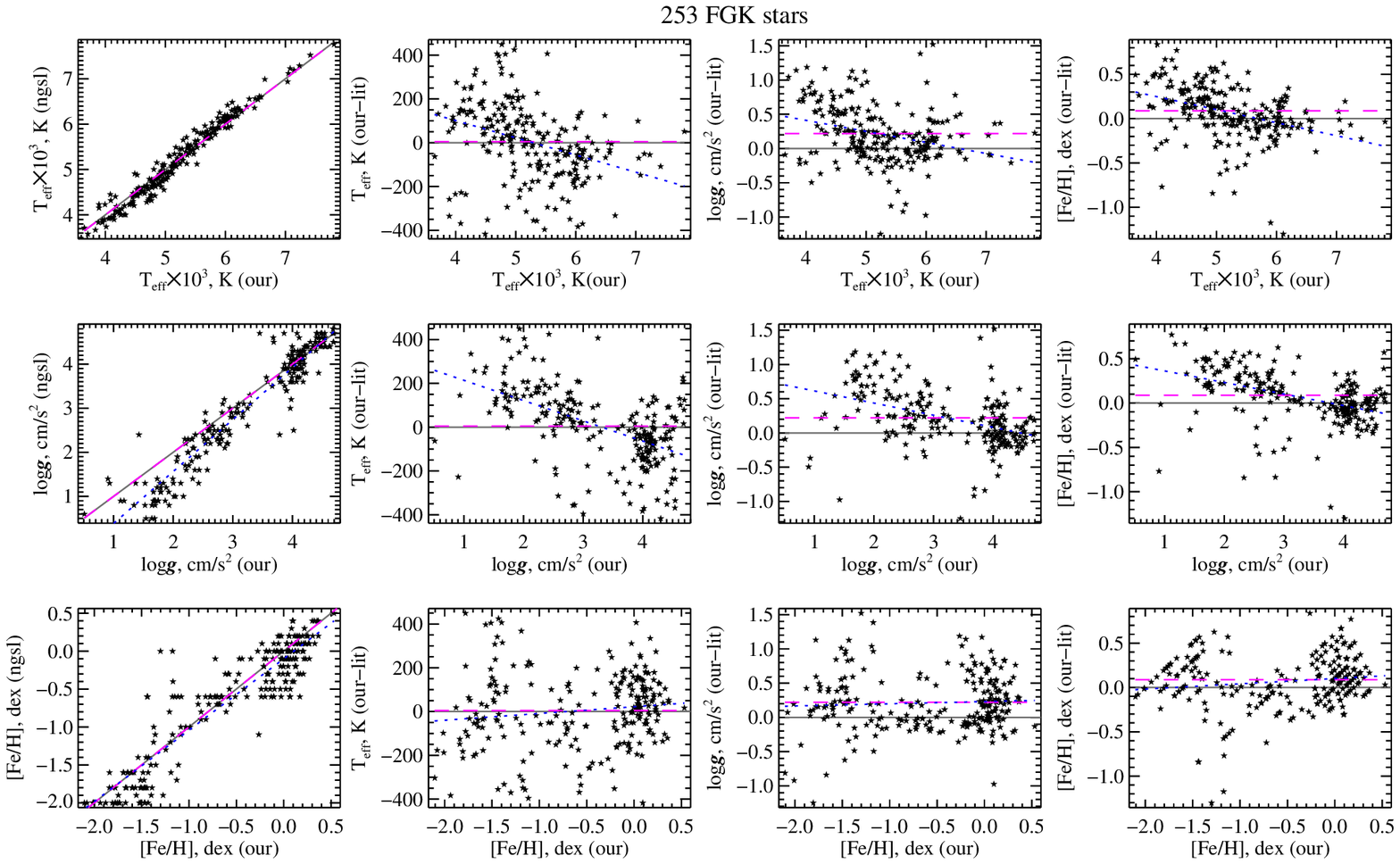}
\caption{
Comparison of the original stellar atmospheric parameters published for the
NGSL with those derived here using \ulyss\ and the MILES interpolator. The
median value of the residuals are plotted in magenta-dashed lines, while the
fit to the residuals are plotted as blue-dotted lines.
The one-to-one relation is shown in black.
The upper three rows of panels show the results for the OBA stars. The remaining
sets of panels show the results for the FGK stars. For each set of stellar
spectral types we show in the first column of panels the comparison of the
temperature, gravity, and metallicity values from top to bottom.
In the last three columns of panels we show the residuals in temperature,
gravity, and metallicty (from left to right) as a function of
temperature, gravity, and metallicity (from top to bottom).
}
\label{fig:compNGSL}
\end{figure*}

\begin{figure*}
\label{fig:compMILES}
\includegraphics[width=0.99\textwidth]{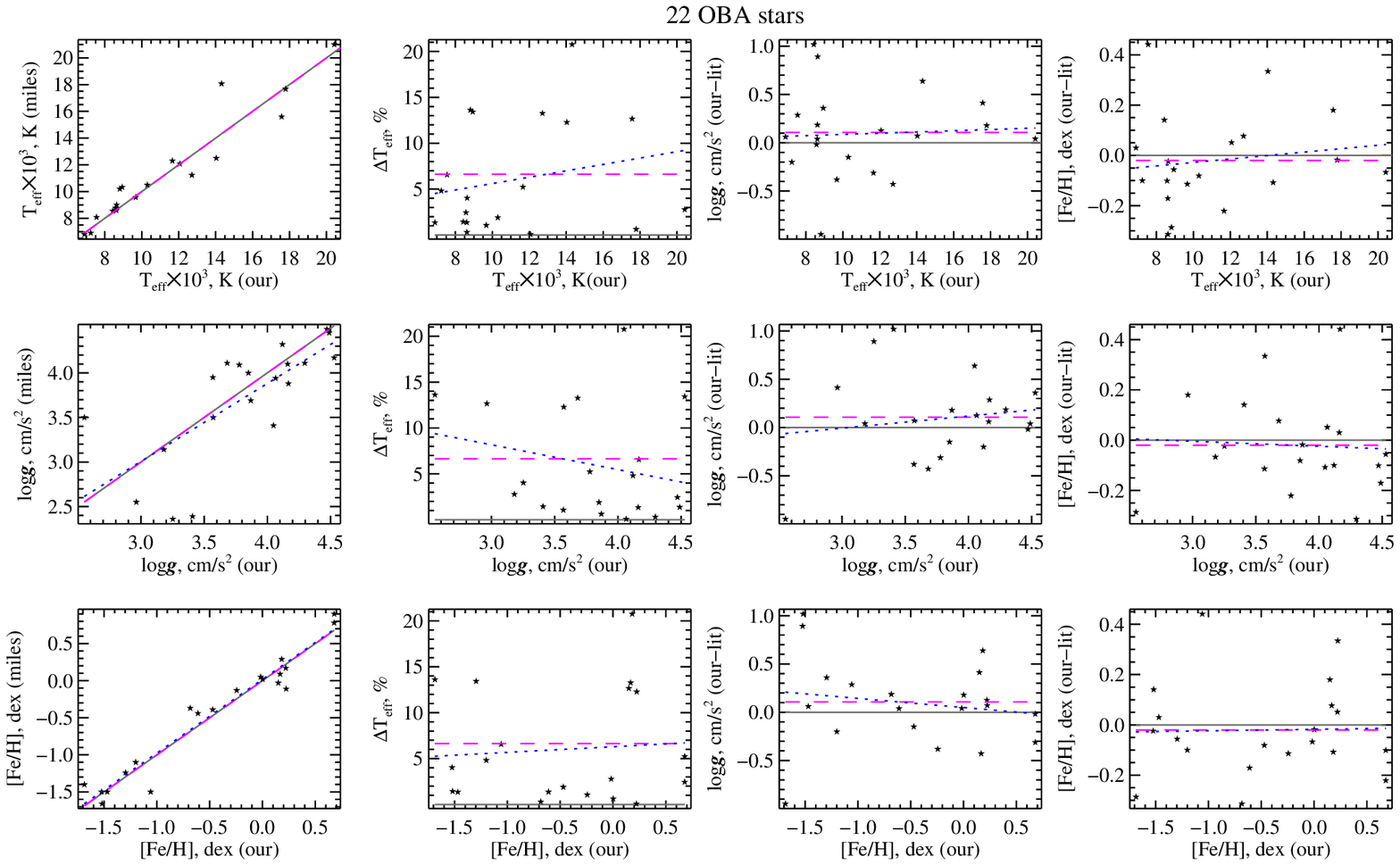}
\includegraphics[width=0.99\textwidth]{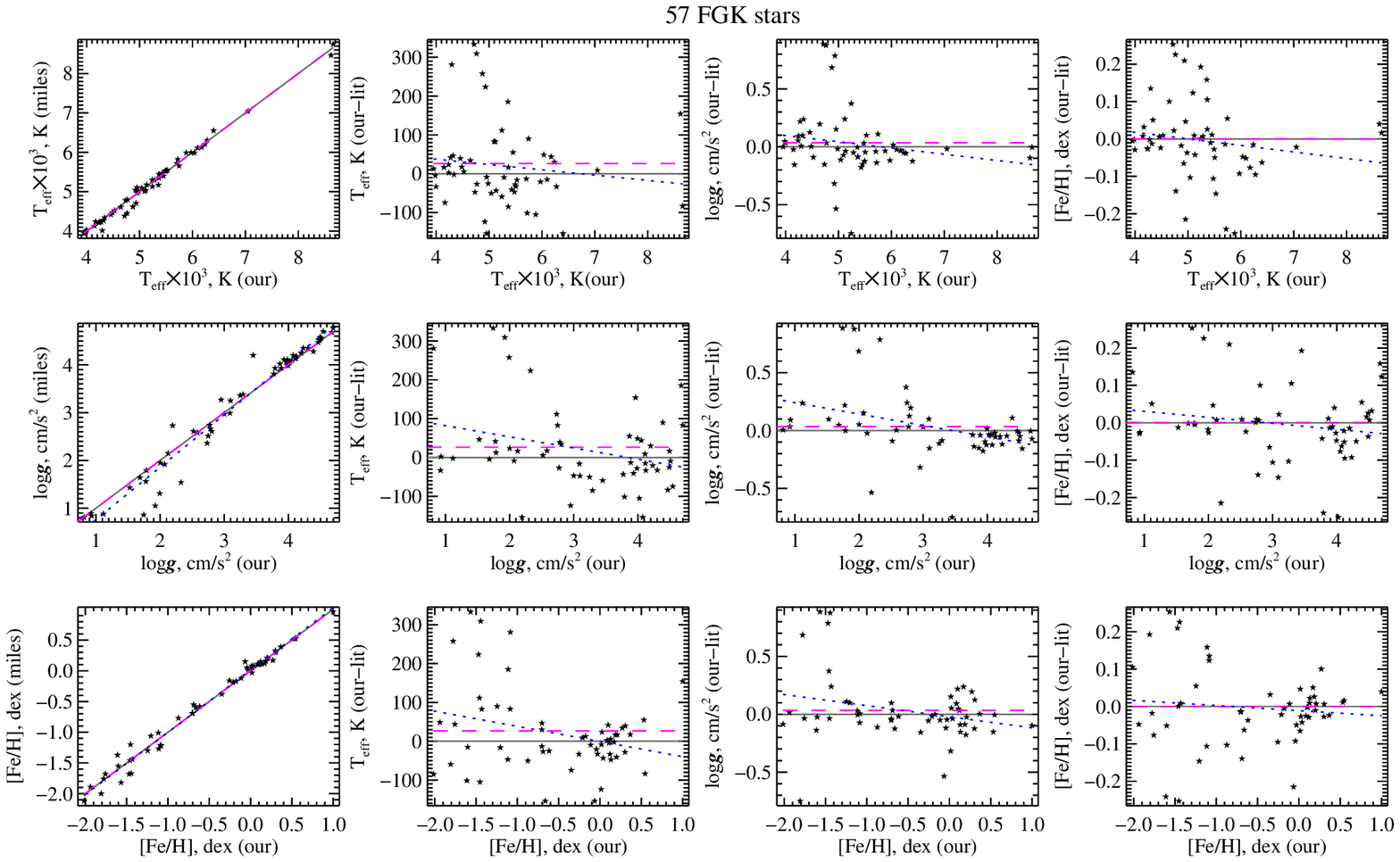}
\caption{Comparison of the stellar parameters derived in this work
  with those from \cite{miles_params} on the basis of the stars in
  common between these two libraries. Points and line types as in
  Fig.\,\ref{fig:compNGSL}.}
\end{figure*}

\begin{figure*}
\label{fig:compCFLIB}
\includegraphics[width=0.99\textwidth]{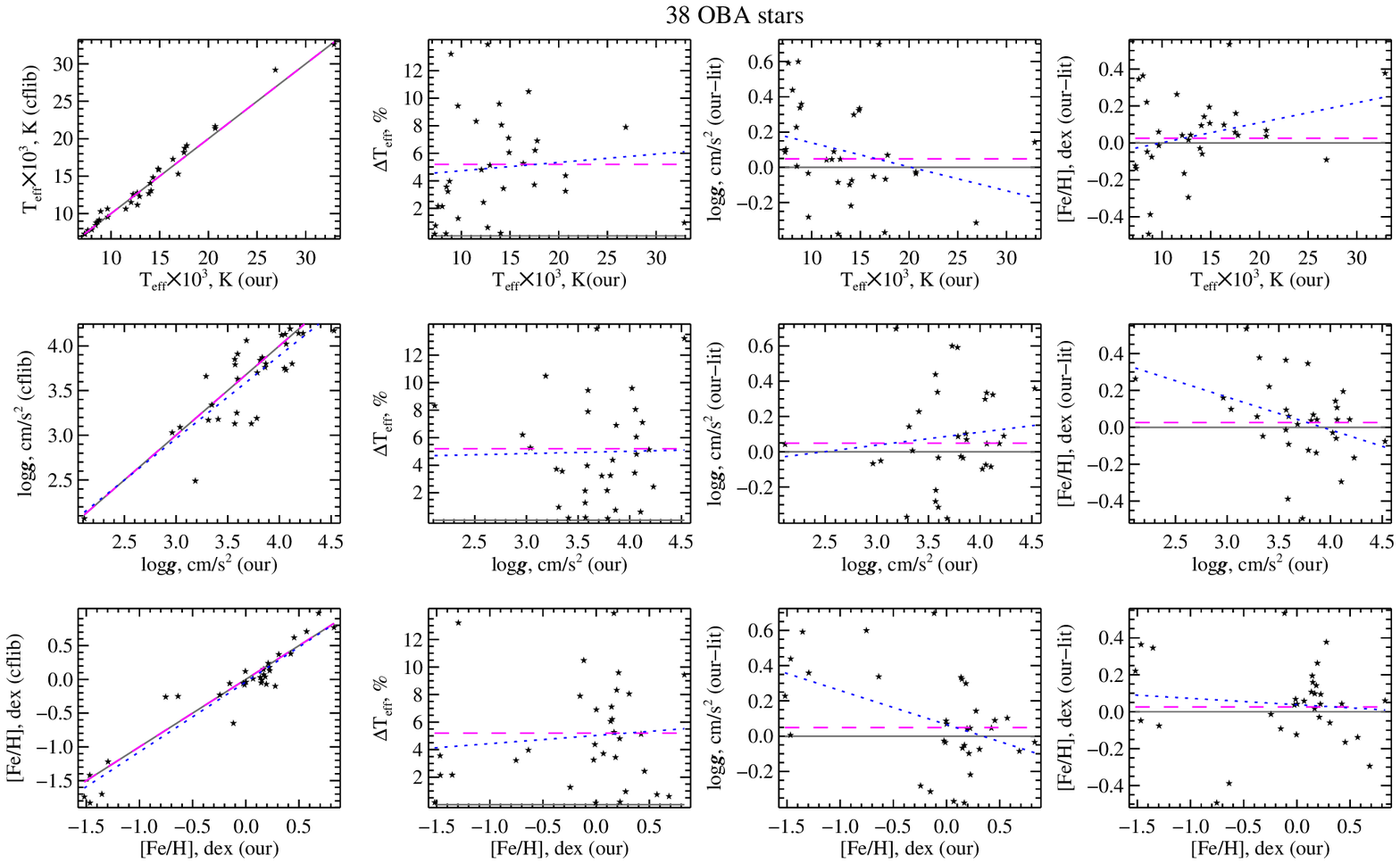}
\includegraphics[width=0.99\textwidth]{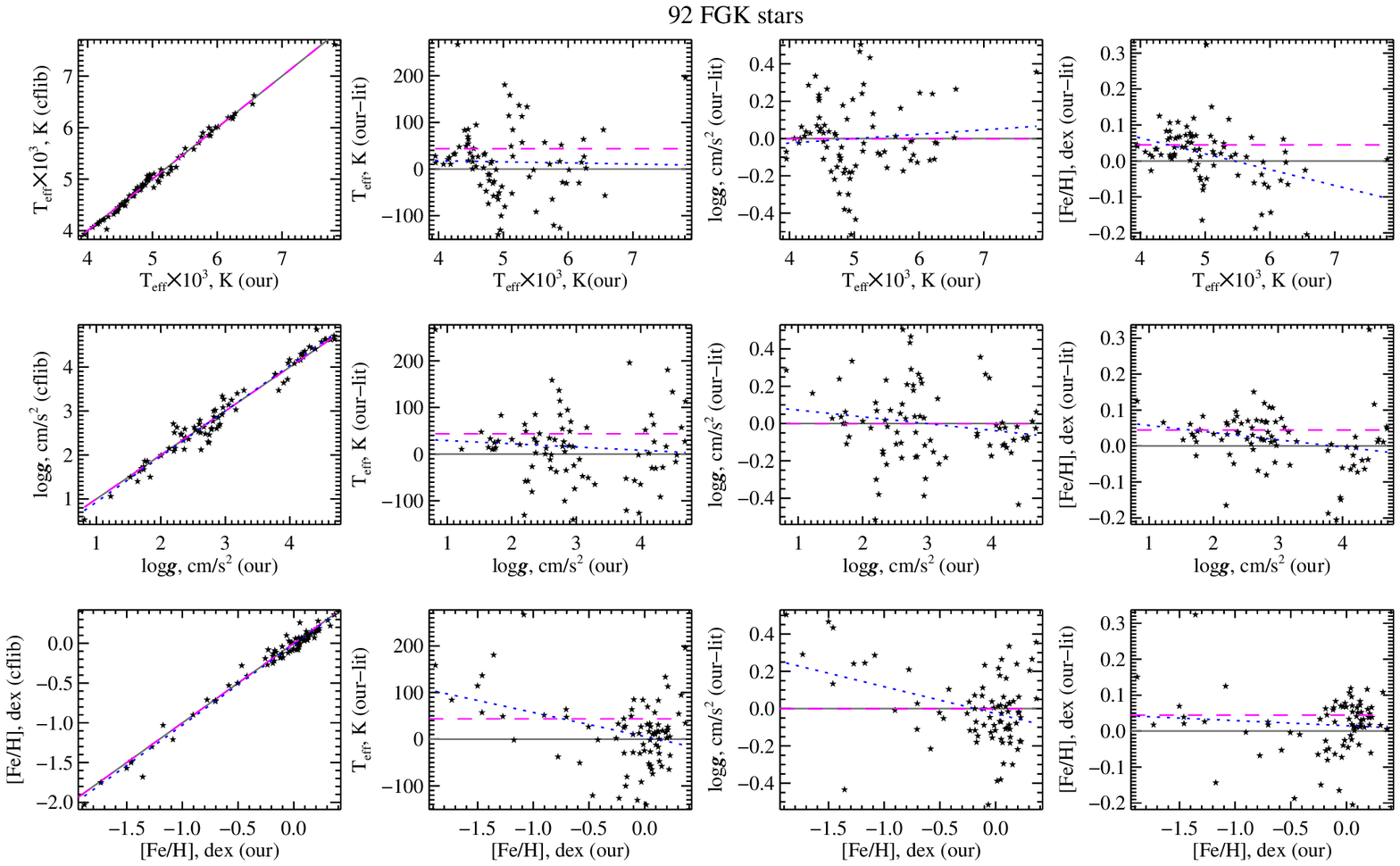}
\caption{Comparison of the stellar parameters derived in this work
  with those from \cite{wu2011} on the basis of the stars in
  common between these two libraries. Points and line types as in
  Fig.\,\ref{fig:compNGSL}.}
\end{figure*}

\begin{figure*}
\label{fig:compPASTEL}
\includegraphics[width=0.99\textwidth]{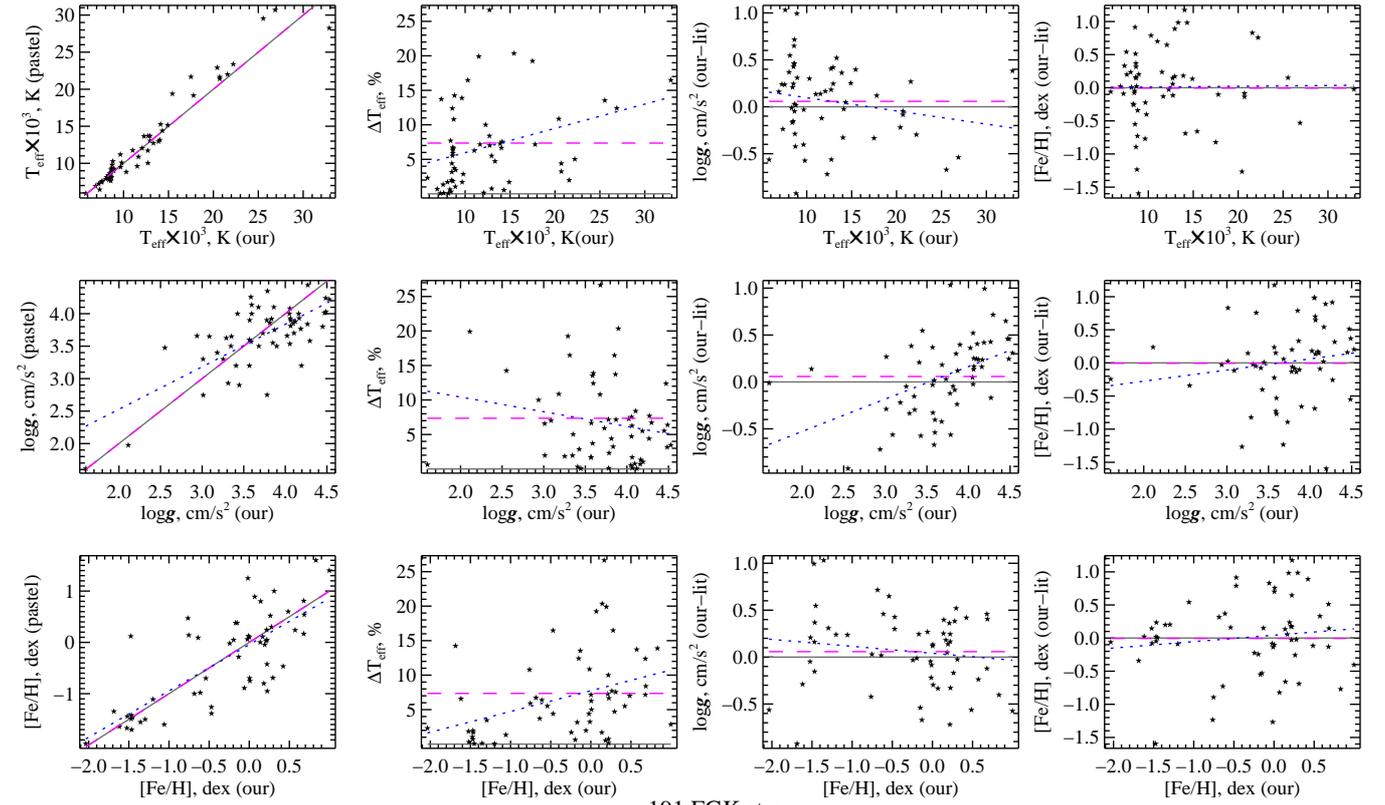}
\includegraphics[width=0.99\textwidth]{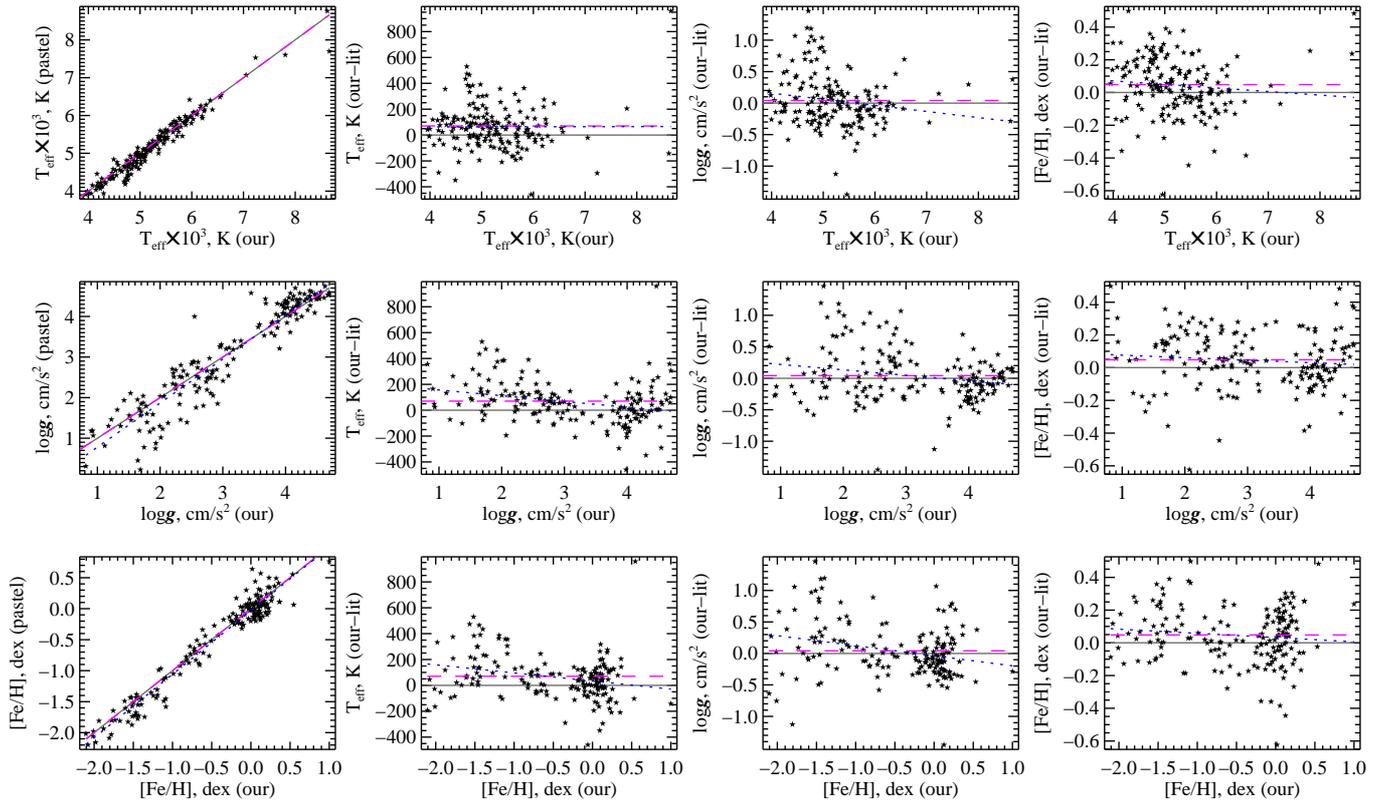}
\caption{Comparison of the stellar parameters derived in this work
  with those from Pastel database on the basis of the stars in
  common between these two libraries. Points and line types as in
  Fig.\,\ref{fig:compNGSL}.}
\end{figure*}

\end{appendix}

\end{document}